\documentclass{article}
\pdfoutput=1

\usepackage{umr-V2}
\usepackage[T1]{fontenc}
\usepackage{epic,eepic,amsmath,latexsym,amssymb,color}
\usepackage{ifthen,graphics,epsfig}
\usepackage[english]{babel}
\usepackage{times}
\usepackage{float}
\usepackage{xspace}

\RRNo{2027}

\RRdate{January 2015}

\RRauthor{
Zohir Bouzid\thanks{ASAP :
\'equipe commune \`a l'IRISA (Universit\'e de Rennes 1) et Inria}
Michel Raynal\thanks{Institut Universitaire de France}\thanks{ASAP :
\'equipe commune \`a  l'IRISA (Universit\'e de Rennes 1) et Inria}
 Pierre Sutra\thanks{University of Neuch\^atel, Switzerland}
}
\authorhead{Z. Bouzid, M. Raynal \& P. Sutra}

\RRetitle{Anonymous Obstruction-free $(n,k)$-Set Agreement\\
           with $n-k+1$  Atomic Read/Write Registers}
\RRtitle{Accord $k$-ensembliste asynchrone et anonyme avec  $(n-k+1)$
         registres atomiques}
\titlehead{Obstruction-free $(n,k)$-Set Agreement with $(n-k+1)$ Registers}
\RRresume{Cet article pr\'esente un algorithme asynchrone
qui r\'esoud l'accord $k$-ensenbliste dans un syst\`eme
de $n$ processus asynchrones et anonymes communiquant
via $(n-k+1)$  registres atomiques du type lire/\'ecrire,
et dans lequel  un nombre quelconque d'entre eux peut s'arr\^er
de f\c{c}on inopin\'ee (crash failure).
La propri\'et\'e de vivacit\'e garantie par l'algorithme est appel\'ee
``{\em obstruction-freedom}''.}
\RRabstract{
The $k$-set agreement problem is a generalization of the consensus problem.
Namely, assuming  each process proposes a value, each non-faulty process
has to decide a value such that each decided value was proposed,
and no more than $k$ different values are decided.
This is a hard problem in the sense that
it cannot be solved in asynchronous systems as soon as $k$ or more processes
may crash. One way to circumvent this impossibility consists in weakening
its termination property, requiring that a process terminates (decides)
only if it executes alone during a long enough period. This is the well-known
obstruction-freedom progress condition.

Considering a system of $n$ {\it anonymous asynchronous} processes, which
communicate through atomic  {\it read/write registers only}, and where
{\it any number of processes may crash}, this paper addresses and solves
the challenging open problem  of designing an  obstruction-free
$k$-set agreement algorithm with $(n-k+1)$ atomic registers only.
From a shared memory cost point of view, this algorithm is the best
algorithm known so far, thereby establishing a new upper bound on the number
of registers  needed to solve the problem (its gain is $(n-k)$
with respect to the previous upper bound). The algorithm is then extended to
address the repeated version of $(n,k)$-set agreement.  As it is optimal
in the number of atomic read/write registers,  this algorithm
closes the gap on previously established lower/upper bounds
for both the anonymous and non-anonymous versions of the repeated
$(n,k)$-set agreement problem. Finally,
for $1 \leq x\leq k<n$, a generalization suited to $x$-obstruction-freedom
is also described, which requires $(n-k+x)$ atomic registers only.
 }
\RRmotcle{
Accord $k$-ensembliste,
Borne de complexit\'e,
Consensus,
syst\`eme asynchrone,
syst\`eme anonyme,
registres atomiques read/write,
crash de processus,
calcul distribu\'e,
tol\'erance aux fautes.
}
\RRkeyword{
Anonymous processes,
Asynchronous system,
Atomic read/write register,
Bounded number of registers,
Consensus,
Distributed algorithm,
Distributed computability,
Fault-tolerance,
$k$-Set agreement,
Obstruction-freedom,
Process crash,
Repeated $k$-set agreement,
Upper bound.
}

\newlength {\squarewidth}

\newcounter{linecounter}
\newcommand{\linenumbering}{\ifthenelse{\value{linecounter}<10}{(0\arabic{linecounter})}{(\arabic{linecounter})}}
\renewcommand{\line}[1]{\refstepcounter{linecounter}\label{#1}\linenumbering}
\newcommand{\resetline}[1]{\setcounter{linecounter}{0}#1}
\renewcommand{\thelinecounter}{\ifnum \value{linecounter} > 9\else 0\fi \arabic{linecounter}}

\newcommand{\Xomit}[1]{}

\newcommand{\true}{\mathit{\tt true}}
\newcommand{\false}{\mathit{\tt false}}
\newcommand{\up}{\mathit{\tt up}}
\newcommand{\down}{\mathit{\tt down}}
\newcommand{\conflict}{\mathit{conflict}}
\newcommand{\conflictf}{\mathit{conf{\ell}ict}}
\newcommand{\conflictOne}{\mathit{conf{\ell}ict}1}
\newcommand{\conflictTwo}{\mathit{conf{\ell}ict}2}
\newcommand{\cfl}{\mathit{cf{\ell}}}
\newcommand{\val}{\mathit{va{\ell}}}
\newcommand{\level}{\mathit{{\ell}eve{\ell}}}
\newcommand{\lvl}{\mathit{{\ell}v{\ell}}}

\newcommand{\sn}{{\color{blue}\mathit{sn}}}
\newcommand{\sni}{{\color{blue}\mathit{sn_i}}}

\newcommand{\dec}{{\color{blue}\mathit{dec}}}
\newcommand{\dcd}{{\color{blue}\mathit{dcd}}}
\newcommand{\dcdi}{{\color{blue}\mathit{dcd_i}}}

\newcommand{\SM}{\mathit{SM}}
\newcommand{\REG}{\mathit{REG}}
\newcommand{\DEC}{\mathit{DEC}}
\newcommand{\Omit}[1]{}

\newcommand{\homog}{\mathcal{H}}

\newcommand{\asign}{\ensuremath{\sqsupset}}
\newcommand{\asigneq}{\ensuremath{\sqsupseteq}}
\newcommand{\asigninv}{\ensuremath{\sqsubset}}
\newcommand{\asigninveq}{\ensuremath{\sqsubseteq}}
\newcommand{\arel}[2]{#1 \asign  #2}

\newcommand{\wrt}{\mathcal{W}}
\newcommand{\elt}{\mathcal{E}}

\newtheorem{definition}{Definition}
\newtheorem{theorem}{Theorem}
\newtheorem{lemma}{Lemma}

\newtheorem{notation}{Notation}

\newcommand{\toto}{xxx}
\newenvironment{proofT}{\noindent{\bf Proof }}
{\hspace*{\fill}$\Box_{Theorem~\ref{\toto}}$\par\vspace{3mm}}
\newenvironment{proofL}{\noindent{\bf Proof }}
{\hspace*{\fill}$\Box_{Lemma~\ref{\toto}}$\par\vspace{3mm}}

\newenvironment{lemma-repeat}[1]{\begin{trivlist}
\item[\hspace{\labelsep}{\bf\noindent Lemma~\ref{#1} }]}%
{\end{trivlist}}

\newenvironment{theorem-repeat}[1]{\begin{trivlist}
\item[\hspace{\labelsep}{\bf\noindent Theorem~\ref{#1} }]}
{\end{trivlist}}
\begin{document}
\makeRR

%=========================================================================
\section{Introduction}

\paragraph{A first challenge: cope with multi-writer atomic registers}
Pioneering works (such as~\cite{L77,P83}) have shown that
processes have to  cope not only with  finite asynchrony
(finite but arbitrary process  speed) but also with
infinite asynchrony (process crash failures), a context in which
mutex-based synchronization  mechanisms  become useless.
This approach has promoted the  design of concurrent algorithms
as a  central topic of {\it fault-tolerant} distributed computing.
See for example Herlihy's  seminal paper~\cite{H91},
or recent textbooks such  as~\cite{HS08,R13,T06}.

When  processes may communicate with {\it Single-Writer Multi-Reader} (SWMR)
atomic registers, a concurrent algorithm usually associates an SWMR register
with each process. This type of registers allows any process to give
information to all the other processes by writing in its own register,
and obtain information from them  by reading their SWMR registers.
The classical snapshot algorithm  introduced in~\cite{AADGMS93} is a
well-known  example of use of such atomic registers.

When  processes communicate with {\it Multi-Writer Multi-Reader} (MWMR)
atomic  registers,  the situation is different. As any process can write
any register, the previous association is no longer given for free.
An approach to  cope with such registers consists in emulating SWMR
registers on top of MWMR registers, and then benefit from existing
SWMR-based algorithms. It is shown in~\cite{DFGL13,DFGR15} that, in
a system of $n$ processes, (a) $(2n-1)$ MWMR atomic registers are needed  to
``wait-free'' simulate one SWMR atomic register, and (b) only $n$
 MWMR atomic registers are needed if the simulation is required to
be only ``non-blocking''\footnote{``Wait-free'' means that any
read or write invocation on the SWMR register that is built must terminate
if the invoking process does not crash~\cite{H91}.
``Non-blocking'' means that at least one process that does not crash
returns from all its read and  write invocations~\cite{HW90}.}.

This simulation approach becomes irrelevant  if  the underlying system
provides the $n$ processes with less than $n$ atomic MWMR registers.
So, we focus here on what we name {\it genuine} concurrent algorithms,
where ``genuine'' means ``without simulating  SWMR registers on top of
MWMR registers''.  An important question is then ``Given a problem, how
many MWMR atomic registers are needed to solve it with a genuine algorithm?''
Unfortunately, as stressed in~\cite{DFGR13}, the design of genuine algorithms
based on  MWMR atomic registers is still in its infancy,  and sometimes
resembles ``black art'' in the sense that their underlying  intuition is
difficult to capture and formulate.

\paragraph{A second challenge: cope with anonymous processes}
In some algorithms based on MWMR atomic registers, a process is required
to write a pair made up of the data value it wants to write, plus  control
values, those including its  identity. This is for example the case of snapshot
algorithms based on MWMR atomic registers~\cite{R13}.

So, a  second question that comes to mind is:
``Is it possible to solve a given problem with MWMR atomic registers and
{\it anonymous} processes; moreover, if the answer is ``yes'', how many
registers are needed?''
To be more precise, let us recall that, in an anonymous system, processes
have no identity,  have the same code, and the same initialization of
their local variables. It is common to remind that, due to privacy
motivations,  anonymous systems  are becoming more and more important.

\paragraph{Consensus and  $k$-set agreement}
The paper considers the  $k$-set agreement problem in a system of $n$
processes. This problem, introduced in~\cite{C93}, and  denoted $(n,k)$-set
agreement  in the following, is a generalization of  consensus,
which corresponds to the instance where $k=1$.
Assuming each participating process  proposes
a value, each non-faulty process must decide a value (termination),
which was proposed by some process (validity), and at most $k$ different
values can be decided (agreement).

\paragraph{Impossibility results and the obstruction-freedom progress condition}
It is well-known that it is impossible to design a deterministic
wait-free consensus algorithm
in asynchronous systems prone to even a single crash failure,
be the underlying communication medium  an asynchronous  send/receive
network~\cite{FLP85}, or a set of read/write atomic registers~\cite{LA87}.
It is also shown in~\cite{BG93,HS99,SZ00} that, if $k$ or more processes may
crash, there is no deterministic wait-free read/write algorithm that can
solve $(n,k)$-set agreement.

As we are interested in the computing power of {\it pure read/write}
asynchronous systems, we  want to neither enrich the underlying system
with additional power such as synchrony assumptions, random numbers, or
failure detectors, nor impose constraints restricting the input vector
collectively proposed by the processes.
So, we consider here a progress condition weaker than wait-freedom,
named {\it obstruction-freedom}~\cite{HLM03}.
In the consensus or $(n,k)$-set agreement context, obstruction-freedom
requires a process to decide a value only
if it executes solo during  a ``long enough period''  (which means that,
during this period,  it is not bothered by other processes).
An in-depth study of complexity issues of obstruction-free algorithms
is presented in~\cite{AGHK09}.

Several obstruction-free consensus algorithms suited to non-anonymous
systems have been proposed (e.g.,~\cite{DFGR13,ELMS05} to cite a few).
When considering anonymous systems,  the obstruction-free algorithm presented
in~\cite{GR07} requires $(8n+2)$ MWMR atomic registers to solve consensus,
and the obstruction-free algorithms described in~\cite{DFGR13,DFKR15}
solve $(n,k)$-set  agreement with $2(n-k)+1$ underlying MWMR atomic registers.

\paragraph{Motivation and content of the paper}
This paper presents  a {\it  genuine obstruction-free} algorithm solving the
$(n,k)$-set agreement problem in an {\it asynchronous anonymous read/write}
system where any number of processes may crash. This algorithm
(called {\it base algorithm} in the following) requires
$(n-k+1)$ MWMR atomic registers (i.e., exactly $n$ registers when one is
interested in the consensus problem).

It is shown in~\cite{EHS98} that $\Omega(\sqrt{n})$
MWMR atomic registers is a lower bound for obstruction-free consensus.
This lower bound has recently been generalized to
$\Omega(\sqrt{\frac{n}{k}-2})$ for  $(n,k)$-set  agreement in anonymous
systems~\cite{DFKR15}.
On another hand, and as already pointed out, the best obstruction-free
$(n,k)$-set agreement algorithm known so  far requires $2(n-k)+1$ MWMR
registers~\cite{DFGR13,DFKR15}.
Hence, the base algorithm  proposed in this paper
provides us with a gain of $2(n-k)+1-(n-k+1)=(n-k)$ MWMR atomic registers.

In the  {\it repeated} version of the  $(n,k)$-set  agreement problem,
the processes participate in a sequence of $(n,k)$-set  agreement instances.
It is shown in~\cite{DFKR15} that $(n-k+1)$ atomic registers
are necessary to solve repeated $(n,k)$-set agreement, be the system
anonymous or non-anonymous. The present paper shows that a simple
modification of the base  obstruction-free $(n,k)$-set agreement
algorithm solves the {\it repeated} $(n,k)$-set agreement problem
without requiring additional atomic registers. It follows that,
as this algorithm requires  $(n-k+1)$ atomic registers, it is optimal,
which closes the gap on previous proposed upper bounds for
the repeated $(n,k)$-set agreement problem.

To attain its goal, the
proposed base
algorithm, which is round-based,
follows the execution pattern ``snapshot; local computation; write'',
where the snapshot and the write are on the $(n-k+1)$ MWMR atomic registers.
This  pattern is reminiscent of the one called
``look; compute; move'' introduced in~\cite{FPSW99,SY96}
in the context of  robot algorithms. Interestingly, no process needs to
maintain local information between successive rounds.
In this sense, the algorithm is {\it locally memoryless}.

From a more technical point of view, each atomic register contains a
quadruplet consisting of a round number, two control bits, and a
proposed value (whose size depends only on the application).
The algorithm exploits a partial order on the quadruplets
that are written into MWMR atomic registers. The way each process
computes new quadruplets is the key of the algorithm.
(The extended version for repeated $(n,k)$-set agreement,
requires sixuplets.)

\paragraph{Roadmap}
The paper is composed of~\ref{sec:conclusion} sections.
Section~\ref{sec:model} presents the computing model and definitions used
in the paper. The presentation is done incrementally.
First, Section~\ref{sec:algorithm-description}  presents  the base
obstruction-free algorithm solving consensus.
This algorithm captures the essence of the solution.
It is proved correct in Section~\ref{sec:algorithm-proof}.
Then, Section~\ref{sec:k-set-agreement}  extends this base algorithm
to obtain an anonymous obstruction-free algorithm solving $(n,k)$-set
agreement, and Section~\ref{sec:repeated} addresses the case
where $(n,k)$-set agreement is used repeatedly.
Section~\ref{sec:x-obstruction-freedom} extends
the base algorithm  to the $x$-obstruction-freedom progress condition
(only $(n-k+x)$ registers are then required by the algorithm).
Finally, Section~\ref{sec:conclusion} concludes the paper.

%======================================================================
\section{Computation Model  and Obstruction-free Consensus}
\label{sec:model}

\subsection{Computing Model}

\paragraph{Process model}
The system is composed of $n$ asynchronous processes, denoted $p_1$, ...,
$p_n$. When considering a process $p_i$, the integer  $i$ is called its index.
Indexes are used  to facilitate the exposition from an external observer
point of view. Processes do not have identities and have the very same code.
We assume that they
know the value $n$.

Up to $(n-1)$ processes may crash. A crash is an unexpected halting.
After it has crashed (if it ever does), a process remains crashed forever.
From a terminology point of view, and given an execution,
a {\it faulty} process is a process that crashes, and
a {\it correct} process is a process that does not crash\footnote{No process
knows if it is correct or faulty. This is because,  before crashing, a faulty
process behaves as a correct process.}. \\

Let $\mathbb{T}$ denote the  increasing sequence of time instants
(observable only from an external point of view).
At each instant, a unique process is activated to execute a  step.
A {\it step} consists in a write or a read of an atomic register (access to
the shared memory) possibly followed by a finite number of
internal  operations (on the local variables of the process that issued the
operation).

\paragraph{Communication  model}
In addition to processes, the computing model includes
a communication medium made up of  $m$ atomic
multi-writer/multi-reader (MWMR) atomic registers\footnote{Let us notice
that the anonymity assumption prevents processes from using
single-writer/multi-reader registers.};
the value of $m$ depends on the problem we want to solve.
These registers are encapsulated in an array denoted $\REG[1..m]$.

``{\it  Atomic}'' means that the read and write operations on a register
$\REG[x]$, $1\leq x\leq m$, appear as if they have been executed sequentially,
and this sequence  (a) respects the real-time order of non-concurrent
operations, and (b) is such that  each read returns the value written by
the closest preceding write operation~\cite{L86}.
When considering any concurrent  object defined from a sequential
specification, atomicity is called {\it linearizability}~\cite{HW90}.
More  generally, the sequence of operations is called
a  {\it linearization}, and the time instant at which an operation appears as
being  executed is called its {\it linearization point}.

\paragraph{From atomic registers to a snapshot object}
At the upper layer (where consensus or $(n,k)$-set agreement is solved),
the array
$\REG[1..m]$ is used to define a snapshot object~\cite{AADGMS93}.
This object, denoted $\REG$,  provides the processes with two operations
denoted ${\sf write}()$ and  ${\sf snapshot}()$.

When a process invokes $\REG.{\sf write}(x,v)$ it deposits the value $v$ in
$\REG[x]$. When it  invokes  $\REG.{\sf snapshot}()$ it obtains the value
of the whole array. The snapshot object is atomic (see above),
which means that each  invocation of $\REG.{\sf snapshot}()$
appears as if it executed instantaneously.
Hence, at this observation level, a linearization is a sequence of
write and snapshot  operations.

An anonymous non-blocking (hence obstruction-free) implementation of a
snapshot object is described  in~\cite{GR07}
(for completeness this algorithm is presented in
Appendix~\ref{annex:OB-free-snapshot}).
This implementation does not require additional atomic registers.
In the following we consider that this snapshot abstraction is
supplied by this underlying layer.

%----------------------------------------------------------------------
\subsection{Obstruction-free consensus and obstruction-free
$(n,k)$-set agreement}
\label{sec:OB-consensus-definition}

\paragraph{Obstruction-free consensus}
An obstruction-free consensus object is a one-shot object that
provides each process with a single operation denoted ${\sf propose}()$.
This operation takes a value as  input parameter and returns a value.

``{\it One-shot}'' means that a process invokes ${\sf propose}()$
at most once.
When a process invokes  ${\sf propose}(v)$, we say that it ``proposes $v$''.
When  the invocation of ${\sf propose}()$ returns  value $v$, we say that
the invoking process ``decides $v$''. A process executes ``solo'' when
it keeps on executing while the other processes have stopped their execution
(at any point of their algorithm).
The obstruction-free consensus problem is defined by the following properties
(that is, to be correct,  any  obstruction-free algorithm must satisfy
these properties).
\begin{itemize}
\vspace{-0.1cm}
\item Validity.
If a process decides a value $v$,  this value was proposed by a process.
\vspace{-0.2cm}
\item Agreement.
No two processes decide different values.
\vspace{-0.2cm}
\item OB-termination.
If there is a time after which a  process executes solo,
it decides a value.
\vspace{-0.2cm}
\item SV-termination\footnote{This termination property, which relates
termination to the input values, is not part of the classical definition of the
obstruction-free consensus problem. It is an additional requirement
which demands  termination under specific circumstances that are
independent of the concurrency pattern.}.
If a single value is proposed, all correct processes decide.
\end{itemize}
Validity relates outputs to inputs. Agreement relates the outputs.
Termination states the conditions under which a correct process must decide.
There are two cases. The first is related to obstruction-freedom. The second
one is independent of the concurrency and failure pattern; it is related to
the input value pattern.

%----------------------------------------------------------------------
\paragraph{Obstruction-free $(n,k)$-set agreement}
\label{sec:OB-k-set-agreement-definition}
An obstruction-free $(n,k)$-set agreement object is a one-shot object which has
the same validity, OB-termination, and SV-termination properties as consensus,
and where the agreement property is:
\begin{itemize}
\vspace{-0.1cm}
\item Agreement.
At most $k$ different values are decided.
\end{itemize}
As for consensus,  SV-termination property is a new property strengthening
the classical definition of $k$-set agreement stated in~\cite{C93}.

%======================================================================
\section{Obstruction-free Anonymous Consensus  Algorithm}
\label{sec:algorithm-description}
The algorithm is described in Figure~\ref{fig:algorithm-OB-consensus}.
As indicated in the Introduction, its essence is captured by
the  quadruplets that can be written in the MWMR atomic registers.

\paragraph{Shared memory}
The shared memory is made up of a snapshot object
$\REG$, composed of $m=n$ MWMR atomic registers.  Each of them
contains a quadruplet initialized to $\langle 0, \down, \false, \bot\rangle$.
The meaning of these fields is the following.
\begin{itemize}
\vspace{-0.2cm}
\item The first field, denoted $rd$, is a round number.
\vspace{-0.2cm}
\item The second field, denoted $\lvl$ (level),
has a value in $\{\up,\down\}$, where $\up > \down$.
\vspace{-0.2cm}
\item The third  field, denoted $\cfl$ (conflict), is a Boolean
(init to $\false$). We assume $\true > \false$.
\vspace{-0.2cm}
\item
The last field, denoted $\val$, is initialized to $\bot$,
and then contains always a proposed value. It is assumed that the
set of proposed values is totally ordered, and the
default value $\bot$ is smaller than any of them.
\end{itemize}
When considering lexicographical ordering, it is easy to see that all possible
quadruplets  $\langle rd, \lvl, \cfl, \val \rangle$ are totally ordered.
This total order, and its reflexive version, are  denoted "$<$''
and ``$\leq$'', respectively.

%========================================================================
\begin{figure}[ht]
\centering{
\fbox{
\begin{minipage}[t]{150mm}
\footnotesize
\renewcommand{\baselinestretch}{2.5}
\resetline
\begin{tabbing}
aaaaa\=aa\=aaa\=aaa\=aaaaa\=aaaaa\=aaaaaaaaaaaaaa\=aaaaa\=\kill %~\\

{\bf function}  ${\sf sup}(T)$ {\bf is}
 ~~\%  $T$ is a set of quadruplets  \% \\

(S1) \> {\bf let}  $\langle r,\level,-, v \rangle$
            {\bf be}  ${\sf max}(T)$;
 ~~~~~~~~~~~~~~~~~~~~\% lexicographical order \%  \\

(S2) \> {\bf let} $vals(T)$  {\bf be}
            $\{ w ~|~ \exists \langle r,-,-,w \rangle\in T\}$; \\

(S3) \> {\bf let} $\conflictOne(T)$  {\bf be}
             $\exists ~\langle r,-,\true, -\rangle \in T$;
              ~~~~~~\% conflict inherited \% \\

(S4) \> {\bf let} $\conflictTwo(T)$  {\bf be} $|vals(T)| >1$;
             ~~~~~~~~~~~~~~~~\% conflict discovered \%  \\

(S5) \> {\bf let}
$\conflictf(T)$ {\bf be} $\conflictOne(T) \vee \conflictTwo(T)$;\\

(S6) \> %{\bf if} $\conflictf(T)$
%                 {\bf then} $\level \gets \conflict$ {\bf end if};\\
%(S7)  \>
${\sf return}
             \big(\langle r,\level,\conflictf(T), v\rangle\big)$.

\end{tabbing}
\normalsize
\end{minipage}
}
\caption{The function  ${\sf sup}()$}
\label{fig:algorithm-sup}
}
\end{figure}

%=======================================================================
\paragraph{The notion of a conflict and the function  ${\sf sup}()$}
The function  ${\sf sup}()$,   defined in Figure~\ref{fig:algorithm-sup},
plays a  central role  in the obstruction-free $(n,k)$-agreement algorithm.
It takes a non-empty set of quadruplets $T$ as input parameter, and
returns a quadruplet, which is the supremum of $T$,
defined as follows.

Let $\langle r,\level, -, v \rangle$ be the maximal element of $T$
according to lexicographical ordering  (line S1), and  $vals(T)$
the values in the quadruplets of $T$ associated with the maximal round number
$r$ (line S2).
The set $T$ is {\it conflicting} if one of the two following cases occurs
(line S5).
\begin{itemize}
\vspace{-0.2cm}
\item
There is a quadruplet $X=\langle r,-,\true, -\rangle$ in $T$ (line S3).
In this case, there is a quadruplet $X\in T$ whose round number is the highest
($X.rd=r$), and whose conflict field $X.\lvl=\true$.
We then say that the conflict is ``inherited''.
\vspace{-0.2cm}
\item
There are at least two quadruplets $X$ and $Y$ in $T$,
that have the highest round number in $T$  (i.e., $X.rd=Y.rd=r$),
and contain different values  (i.e., $X.\val\neq Y.\val$) (lines S2 and S4).
In this case we say  say that the conflict is ``discovered''.
\end{itemize}

The function ${\sf sup}(T)$  first  checks  if $T$ is conflicting
(lines S2-S5). Then it  returns at line S6 the quadruplet
$\langle r,\level, \conflictf(T), v\rangle$, where $\conflictf(T)$ indicates
if the input set $T$  is conflicting (line S5). Let us notice that,
since $\true>\false$, the quadruplet returned by
${\sf sup}(T)$ is always greater than, or equal to, the greatest element in
$T$, i.e., ${\sf sup}(T) \geq {\sf max}(T)$.

%========================================================================
\begin{figure}[ht]
\centering{
\fbox{
\begin{minipage}[t]{150mm}
\footnotesize
\renewcommand{\baselinestretch}{2.5}
\resetline
\begin{tabbing}
aaaaa\=aa\=aaa\=aaa\=aaaaa\=aaaaa\=aaaaaaaaaaaaaa\=aaaaa\=\kill %~\\

{\bf operation} ${\sf propose}(v_i)$ {\bf is} \\

\line{Z-01} \> {\bf repeat forever} \\

\line{Z-02} \>\> $view \leftarrow \REG.{\sf snapshot}();$\\

\line{Z-03} \> \>
   {\bf case} \= $(\forall x: view[x]=\langle r,\up,\false,\val\rangle
                          ~~~~~~~\text{where }r > 0)$
                          ~\={\bf then} ${\sf return}(\val$)\\

\line{Z-04} \>\>\>$(\forall x: view[x]=\langle r,\down,\false,\val\rangle
            ~~~\text{where }r > 0)$ \>
 {\bf then} $\REG.{\sf write}(1, \langle r+1,\up,\false, \val\rangle)$\\

\line{Z-05} \>\>\> $(\forall x: view[x]=\langle r,\level,\true, \val\rangle
               ~~~~\text{where }r > 0)$  \>

 {\bf then} \=
    $\REG.{\sf write}(1, \langle r+1,\down,\false,  \val\rangle)$;\\

%\line{Z-06} \>\>\>\>\>
%    $\REG.{\sf write}(1, \langle r+1,\down,\false,  w\rangle)$\\

\line{Z-06} \>\>\> {\bf otherwise} \=
{\bf let}   $\langle r,\level,\cfl,\val \rangle \leftarrow
     {\sf sup}(view[1],\cdots,view[n], \langle 1,\down, \false, v_i\rangle)$;\\

\line{Z-07}  \>\> \>\>
    $x \gets$
    smallest index such that $view[x] \neq \langle r,\level,\cfl,\val \rangle $;\\

\line{Z-08}  \>\> \>\>
$\REG.{\sf write}(x, \langle r,\level, \cfl,\val\rangle)$\\

\line{Z-09} \>\> {\bf end case}\\

\line{Z-10} {\bf end repeat}.

\end{tabbing}
\normalsize
\end{minipage}
}
\caption{Anonymous obstruction-free Consensus}
\label{fig:algorithm-OB-consensus}
}
\end{figure}
%=======================================================================

\paragraph{The algorithm}
The algorithm is pretty simple. It consists in an appropriate management
of the snapshot object $\REG$, so that the $n$  quadruplets it contains
(a) never allow validity and agreement to be violated,  and
(b) eventually allow termination under good circumstances (which occur when
obstruction-freedom is satisfied or when a single value is proposed).

When a process $p_i$ invokes ${\sf proposes}(v_i)$, it enters a loop
that it will exit  at line~\ref{Z-03} (if it terminates), by executing
the statement ${\sf return}(\val)$, where $\val$ is the value it decides.

After entering the loop a process issues first a snapshot, and assigns
the returned array to its local variable $view[1..n]$ (line~\ref{Z-01}).
Then,  there are two main cases  according to the value of  $view$.
\begin{itemize}
\vspace{-0.2cm}
\item Case 1 (lines~\ref{Z-03}-\ref{Z-05}).
All entries of  $view_i$ contain the same quadruplet
$\langle r,\level, \conflict val\rangle$, and  $r > 0$. \\
There are three sub-cases.
\begin{itemize}
\vspace{-0.2cm}
\item  Case 1.1.
If the level is $\up$ and the conflict is $\false$,
the invoking process decides the value $\val$ (line~\ref{Z-03}).
\vspace{-0.1cm}
\item  Case 1.2.
If the level is $\down$  and the conflict field is $\false$,
the invoking process
decides the value $\val$ (line~\ref{Z-03}). is $\false$, process $p_i$
enters the next round by writing $\langle r+1,\up, \false, val\rangle$
in the first entry of $\REG$ (line~\ref{Z-04}).
\vspace{-0.1cm}
\item  Case 1.3. If there is a conflict,  $p_i$
enters the next round by writing $\langle r+1,\down, \false, val\rangle$
in the first entry of $\REG$ (line~\ref{Z-05}).
\end{itemize}
\vspace{-0.2cm}
\item Case 2 (lines~\ref{Z-06}-\ref{Z-08}).
Not all entries of  $view_i$ are equal or
one of them contains  $\langle 0,-,-,- \rangle$.\\
In this case, process
$p_i$ calls the  internal function
${\sf sup}(view[1],\cdots,view[n],\langle 1,\down, \false, v_i\rangle)$
(line~\ref{Z-06}), which returns  a quadruplet $X$ that is greater than
all the input quadruplets  or equal to the greatest of them.
As we have seen, this quadruplet $X$  may inherit or discover a conflict.
Moreover, as $\langle 1,\down,\false, v_i\rangle$  is an input parameter
of the function ${\sf sup}()$, $X.\val$  cannot be $\bot$.

Let us notice that, as none of the predicates of lines~\ref{Z-03}-\ref{Z-05}
is satisfied, not all entries of $view[1..n]$ can be equal to the previous
quadruplet $X$.
The invoking process  $p_i$ writes $X$ into $\REG[x]$, where, from its
point of view,  $x$ is the first entry of $\REG$ whose content is different
from $X$ (lines~\ref{Z-07}-\ref{Z-08}).
\end{itemize}

\paragraph{The underlying operational intuition}
To understand the intuition that underlies the algorithm, let us first consider
the very simple case where a single process $p_i$ executes the algorithm.
It obtains from its first invocation of $\REG.{\sf snapshot}()$
(line~\ref{Z-02}) a view $view$ in which all elements are equal to
$\langle 0,\down,\false,\bot\rangle$.
Hence,  $p_i$ executes line~\ref{Z-06},
where  the invocation of ${\sf sup}()$ returns the quadruplet
$\langle 1,\down, \false,v_i\rangle$, which is written into $\REG[1]$
at  line~\ref{Z-08}.  Then, during the second round,
$p_i$ computes a quadruplet with the help of the function ${\sf sup}()$,
which returns $\langle 1,\down, \false,v_i\rangle$,  and writes this
quadruplet  into $\REG[2]$;  etc., until
$p_i$ has written  $\langle 1,\down, \false,v_i\rangle$ in all the atomic
registers of $\REG[1..n]$. When this has been done, $p_i$ obtains
at line~\ref{Z-02} a view all elements of which are equal to
$\langle 1,\down,\false, v_i\rangle$. It consequently executes
line~\ref{Z-04} and writes $\langle 2,\up, \false,v_i\rangle$ in $\REG[1]$.
Then, during the following  executions of the loop body, it writes
$\langle 2,\up, \false, v_i\rangle$ in the other registers of
$\REG$  (line~\ref{Z-08}).
When this is done, $p_i$  obtains a snapshot containing only the quadruplet
$\langle 2,\up,\false, v_i\rangle$. When this occurs,
$p_i$ is directed to execute line~\ref{Z-03} where it decides.

Let us now consider the case where, while $p_i$ is executing, another process
$p_j$ invokes ${\sf propose}(v_j)$ with $v_j=v_i$.
It is easy to see that $p_i$ and $p_j$ collaborate then to
fill in $\REG$ with the same quadruplet $\langle 2,\up, v_i\rangle$.
If $v_j\neq v_i$, depending on the concurrency pattern,
a conflict may occur. For instance, it occurs if  $\REG$ contains both
$\langle 1,\down,\false, v_i\rangle$ and
$\langle 1,\down, \false,v_j\rangle$.
If a  conflict appears, it will be propagated from round to round,
until a  process  executes alone a higher round number.

\paragraph{Remark 1}
Let us notice that no process needs to memorize in its local memory
values that will be used in the next round. Not only the processes are
anonymous, but their code is memoryless (no persistent variables).
The snapshot object $\REG$ constitutes the whole memory of the system.
Hence, as defined in the Introduction, the algorithm is locally memoryless.
In this sense, and from a locality point of view, it has a ``functional''
flavor.

\paragraph{Remark 2}
Let us consider the $n$-bounded concurrency model~\cite{A04,MT13}.
This model is made up of an arbitrary number of processes, but, at any time,
there are at most $n$  processes executing steps. This allows processes to
leave the system and other processes to join it as long as the concurrency
degree does not exceed~$n$.

The previous algorithm works without modification in such a model.
A proposed value is now a value proposed by any of the $N$
processes that participate in the algorithm.  Hence, if
If $N > n$, the number of proposed values can be greater than the
upper bound $n$ on the concurrency degree.
This versatility dimension of the algorithm is a direct
consequence of  the previous  ``locally memoryless'' property.

%======================================================================
%======================================================================
\section{Proof of the Algorithm}
\label{sec:algorithm-proof}
\label{sec:proof}

After a few definitions provided in Section~\ref{sec:proof-definitions},
Section~\ref{sec:proof-partial-order} shows that the relation
``$\asigneq$'' defined on quadruplets is a partial order. This relation is
 central to prove properties of the algorithm.
Such properties are stated and proved in Sections~\ref{sec:proof-algo-order}
and~\ref{sec:exploiting-homo-snapshots}.
Based on these previous properties, Section~\ref{sec:proof-algo}
establishes the correctness of our algorithm.

%-----------------------------------------------------------------------
\subsection{Definitions and notations}
\label{sec:proof-definitions}
Let $\elt$ be a set of quadruplets that can be written in $\REG$.
Given $X \in \elt$,  its four fields are denoted
$X.rd, X.\lvl, X.\cfl$ and $X.\val$, respectively, and
$>$ and $\geq$ refer to the classical lexicographical order on  $\elt$.
Moreover, where appropriate, an array $view[1..n]$ is considered as the set
 $\{view[1], \cdots, view[n]\}$.
% If this array is from process $p_i$, it is denoted $view_i$.

\begin{definition}
\label{def:domination}
let  $X, Y \in \elt$.
\vspace{-0.3cm}
$$\arel{X}{Y}~ \stackrel{\mathit{def}}{=}~
(X >Y)\wedge
[(X.rd > Y.rd)\vee (X.\cfl)
                              \vee(\neg Y.\cfl \wedge X.\val=Y.\val)].$$
\end{definition}

\Xomit{%%%%%%%%%%%%%%%%%%%%%%%%%%%%%%%%%%%%%%%
\noindent
Let $X$ and $Y$ be two distinct triplets. It  follows from the
definitions of $X>Y$ and $\arel{X}{Y}$ that:
\begin{itemize}
\vspace{-0.2cm}
\item $[(X.\val=Y.\val)\vee(X.\cfl=\true)]\Rightarrow
[(X>Y) \Leftrightarrow (\arel{X}{Y})]$.
\vspace{-0.2cm}
\item $(X.\val\neq Y.\val)\Rightarrow
[(\arel{X}{Y})  \Leftrightarrow \big(X>Y\big)\wedge
    \big((X.rd>Y.rd)\vee (X.\cfl=\true)\big)]$.

\end{itemize}

As an example, let us consider $X=\langle r_X,\up,v_X\rangle$ and
$Y=\langle r_Y,\down,v_Y\rangle$. If $r_X>r_Y$, we have both $X>Y$ and
$X \asign Y$.
If $r_X=r_Y$, as $\up > \down$,  we have $X>Y$, and if additionally
$v_X=v_Y$,  we also have  $X \asign Y$.
If $r_X=r_Y$ and $v_X\neq v_Y$, we have neither $X \asign Y$ nor $Y \asign X$
(while we have $X>Y$ or $Y>X$).
}  %%%  end of \Xomit{%%%%%%%%%%%%%%%%%%%%%%%%%%%%%%%%%%%%%%%

At the operational level the algorithm ensures that the quadruplets it
generates are totally ordered by the relation $>$. Differently, the
relation  $\asign$ (which is a partial order on these quadruplets,
see Section~\ref{sec:proof-partial-order})  captures the relevant part of
of this total order, and is consequently the key cornerstone on which
relies the proof of our algorithm.  \\

\noindent
When $\arel{X}{Y}$, we say  ``$X$ {\it strictly dominates} $Y$''.
$X$  {\it dominates} $Y$, denoted $X \asigneq Y$, if $(X \asign Y)$ or $(X=Y)$
holds.
The relations $\asigninv$ and $\asigninveq$ are defined in the natural way.
\begin{definition}
\label{def:homogeneous}
Given a set of quadruplets $T$, we shall say that
$T$ is {\it homogeneous}  when it contains a single element, say $X$.
 We then write it  ``$T$ is $\homog(X)$''.
%It is {\it pseudo-homogeneous}, denoted $\phomog(X)$,
%if there exists $X$ such that
%every
%element of $T$ is equal to $\langle X.rd, X.\lvl, X.\cfl, - \rangle$.
%$T$ is {\it uniform}, denoted $\unif(X)$,
%if either it is  $\homog(X)$  with $X.\cfl=\false$
%or it is $\phomog(X)$ with $X.\cfl=\true$.
\end{definition}
\begin{notation}
The value,  at time $\tau$, of the local variable  $xxx$  of a process $p_i$
is denoted $xxx_i^\tau$. Similarly the value of an atomic  register $\REG[x]$
at time $\tau$ is denoted  $\REG^\tau[x]$, and the value of  $\REG$
at time $\tau$ is denoted  $\REG^\tau$.
\end{notation}
\begin{notation}
Let  $\wrt(x,X)$ denote the writing of a quadruplet $X$ in the register
$\REG[x]$.
\end{notation}
\noindent
\begin{definition}
\label{def:covezring-write}
We say  ``a process $p_j$ {\em covers} $\REG[x]$ at time $\tau$''
when  its next non-local step after time $\tau$ is $\wrt(x,X)$,
where $X$ is the quadruplet which is written.
In this case we also say ``$\wrt(x,X)$ {\em covers} $\REG[x]$ at time $\tau$''
or ``$\REG[x]$  {\em is covered} by $\wrt(x,X)$ at time $\tau$''.
\end{definition}
\noindent Let us notice that if, at time $\tau$,
$p_j$ covers $\REG[x]$, then $\tau$ necessarily lies between the last snapshot
issued by $p_j$ at line~\ref{Z-02} and its planned write $\wrt(x,X)$
that will occur at line~~\ref{Z-04}, \ref{Z-05}, or~\ref{Z-08}.

%--------------------------------------------------------------------%
\subsection{The relation $\asigneq$  is a partial order}
\label{sec:proof-partial-order}

\begin{lemma}
\label{lemma:for-partial-order}
$((X \asign Y \asign Z) \wedge (X.rd=Y.rd=Z.rd))$
$\Rightarrow$ $(X.\cfl \vee (\neg Z.\cfl \wedge X.\val = Z.\val))$.
\end{lemma}

\begin{proofL}
Let us assume that $(\neg~X.\cfl)$ holds, we have to prove $\neg Z.\cfl$
and $X.\val = Z.\val$.
%Let us assume for contradiction that $(\neg~X.\cfl)$.
It then follows from the lemma assumption and the definition of  $\asign$
that we have:
\vspace{-0.2cm}
$$
((X \asign Y)  \wedge (X.rd=Y.rd) \wedge (\neg X.\cfl ))
\Rightarrow
(\neg Y.\cfl \wedge X.\val = Y.\val).
$$
\vspace{-0.2cm}
\noindent
Hence we can use the same argument as above to show that
$(\neg Z.\cfl \wedge Y.\val = Z.\val)$:
$$
((Y \asign Z)  \wedge (Y.rd=Z.rd) \wedge (\neg Y.\cfl))
\Rightarrow (\neg Z.\cfl \wedge Y.\val = Z.\val). $$
\vspace{-0.2cm}
Summarizing we have $(\neg Z.\cfl \wedge X.\val = Z.\val)$.
This proves the claim.
\renewcommand{\toto}{lemma:for-partial-order}
\end{proofL}

\begin{lemma}
\label{lemma:partial-order}
$\asigneq$ is a partial order.
\end{lemma}

\begin{proofL}
To prove  the transitivity property, let us assume that $X \asigneq Y$
and $Y \asigneq Z$. We have to show  that $X \asigneq Z$.
If $X=Y$ or $Y=Z$, the claim follows trivially.
Hence, let us  assume  that $Y$ is neither $X$ nor $Z$.
As  $(X \asign Y) \Rightarrow (X>Y)$,
 $(Y \asign Z) \Rightarrow (Y>Z)$, it follows that $X>Z$.
To prove $X \asign Z$, it remains to show that
$((X.rd > Z.rd) \vee (X.\cfl) \vee (\neg Z.\cfl \wedge X.\val=Z.\val))$.
Let us observe that, due to the definition of $\asign$,
we have $(X \asign Y)~\Rightarrow~\big((X.rd > Y.rd)
           \vee (X.\cfl) \vee (\neg Z.\cfl \wedge  X.\val=Y.\val)\big)$.
There are three cases.
\begin{itemize}
\vspace{-0.2cm}
\item Case $(X.rd > Y.rd)$.
As $Y \asign Z$ we have  $(Y.rd \geq Z.rd)$.
Hence,  $(X.rd > Z.rd)$.
\vspace{-0.2cm}
\item  Case $(X.rd = Y.rd)$ $\wedge$ $(Y.rd > Z.rd)$.
Then, we have  $(X.rd >Z.rd)$.
\vspace{-0.2cm}
\item  Case $(X.rd = Y.rd)$ $\wedge$ $(Y.rd = Z.rd)$.
Then,  Lemma~\ref{lemma:for-partial-order}
$\Rightarrow$ $(X.\cfl \vee (\neg Z.\cfl \wedge X.\val = Z.\val))$.
\end{itemize}
\vspace{-0.2cm}
In each case,  the transitivity property follows.

To prove the  antisymmetry property,
we show that if $X \asign Y$ then $Y \not\asign X$.
Assume for contradiction that  $X \asign Y$ and $Y \asign X$.
It follows that $X > Y$ and $Y>X$, contradiction.
\renewcommand{\toto}{lemma:partial-order}
\end{proofL}

%--------------------------------------------------------------------
\subsection{Extracting the relations $\asign$ and $\asigneq$ from the algorithm}
\label{sec:proof-algo-order}
The definition of  ${\sf sup}()$ appears in Figure~\ref{fig:algorithm-sup}.

\begin{lemma}
\label{lemma:sup-dominates}
Let $T$ be a set  of quadruplets.
For every $X \in T: {\sf sup}(T) \asigneq X$.
\end{lemma}

\begin{proofL}
Let $X \in T$ and $S={\sf sup}(T)$.  We have to  prove that $S \asigneq X$.
Let us first observe that, as  $S={\sf sup}(T) \geq {\sf max}(T) \geq X$,
we have  $S \geq X$. If $S = X$ then the lemma follows immediately.
So let us assume in the following that $S > X$.
There are two cases.
\begin{itemize}
\vspace{-0.2cm}
\item If $S.rd > X.rd$, then $S \asign X$,  and the lemma follows.
\vspace{-0.2cm}
\item  Assume that $S.rd=X.rd$.
We need to show that $(S.\cfl) \vee (\neg X.\cfl \wedge S.\val = X.\val)$.

In the following we prove that $(\neg S.\cfl \Rightarrow \neg X.\cfl)$.
Therefore we need then only to show that $(S.\cfl) \vee (S.\val = X.\val)$.

Let us first prove $(\neg S.\cfl \Rightarrow \neg X.\cfl)$.
We do it by proving the contrapositive $X.\cfl \Rightarrow S.\cfl$.
 If $(X.\cfl)$, we have the following. Since $X.rd=S.rd={\sf sup}(T).rd$,
it follows that the predicate $\conflictOne(T)$  is true, which implies that
$S.\cfl={\sf sup}(T).\cfl$ is also true. Therefore $X.\cfl \Rightarrow S.\cfl$.

Let us now show the second part, i.e.,
either $(S.\cfl)$ or $(S.\val = X.\val)$ holds.
Assume that $(S.\val \neq X.\val)$ and let us prove that $(S.\cfl)$ is true.
Let us observe that, due to the definition of $S={\sf sup}(T)$
(Figure~\ref{fig:algorithm-sup}),
${\sf max}(T).\val = {\sf sup}(T).\val=S.\val$.
But we assumed $S.\val \neq X.\val$. Therefore ${\sf max}(T).\val \neq X.\val$.
This means that there are at least two elements in $T$,
namely $X$ and ${\sf max}(T)$, which are associated with the maximal round
$S.rd$, and which carry distinct values ($X.\val \neq {\sf max}(T).\val$).
Hence, the predicate $\conflictTwo(T)$ is satisfied, and consequently
${\sf sup}(T).\cfl$ is equal to $\true$. Therefore $S= {\sf sup}(T) \asign X$.
\end{itemize}
\vspace{-0.6cm}
\renewcommand{\toto}{lemma:sup-dominates}
\end{proofL}

%--------------------------------

\begin{lemma}
\label{lemma:write-dominates}
If $p_i$ executes $\wrt(-,Y)$ at time $\tau$, then
for every $X \in view_i^{\tau}: Y \asigneq X$.
\end{lemma}

\begin{proofL}
%There are two cases depending on whether $view_i^{\tau}$ is homogeneous or not.
We consider  two cases according to the line at which the write occurs.

\begin{itemize}
\vspace{-0.2cm}
\item  $Y$ is written at
line~\ref{Z-04} or~\ref{Z-05}.
It follows that $Y.rd = ({\sf max}(view_i^\tau).rd)+1$.
Therefore, for every $X \in view_i^{\tau}: Y.rd > X.rd$. Hence $Y \asign X$.
\vspace{-0.2cm}
\item
 $Y$ is written at
line~\ref{Z-08}.
In this case,  due to the invocation of the function ${\sf sup}()$ at
line~\ref{Z-06}, the value $Y$  written by $p_i$ is equal to ${\sf sup}(T)$
where
$T=\{view_i^{\tau}[1],\cdots,view_i^{\tau}[n],\langle 1,\down,
                                              \false, v_i \rangle\}$.
According to Lemma~\ref{lemma:sup-dominates}, it follows that
for every $X \in view_i^{\tau}$ we have $Y={\sf sup}(T) \asigneq X$.
\end{itemize}
\vspace{-0.4cm}
\renewcommand{\toto}{lemma:write-dominates}
\end{proofL}

%---------------------------------------------------------------------
\begin{lemma}
\label{lemma:tool}
Let us assume that no process is covering  $\REG[x]$ at time $\tau$.
For every write $\wrt(-,X)$ that (a) occurs after $\tau$ and (b)
was not covering a register of $\REG$  at time $\tau$, we have
$X \asigneq \REG^{\tau}[x]$.
\end{lemma}

\begin{proofL}
The proof is by contradiction.
Let $p_i$ be the first process that executes a write $\wrt(-,X)$
contradicting  the lemma. This means that  $\wrt(-,X)$ is not
covering a register of $\REG$ at time $\tau$ and $X \not\asigneq \REG^{\tau}[x]$.
Let this write occur at time $\tau_2 > \tau$.
Thus, all writes that take place between $\tau$ and $\tau_2$ comply with the
lemma. We derive a contradiction by showing that $X \asigneq \REG^{\tau}[x]$.

Let $\tau_1 < \tau_2$ be the linearization time of the last snapshot
taken by $p_i$ (line \ref{Z-02}) before executing $\wrt(-,X)$.
Since $\wrt(-,X)$ was not covering a register of $\REG$ at time $\tau$,
the snapshot  preceding this write  was necessarily taken after $\tau$.
That is, $\tau_1 > \tau$, and we have $\tau_2 > \tau_1 > \tau$.

According to Lemma~\ref{lemma:write-dominates}, $X \asigneq view_i^{\tau_2}[x]$.
But since the snapshot returning $view_i^{\tau_2}$ is linearized at $\tau_1$,
it follows that $view_i^{\tau_2}=\REG^{\tau_1}$. Therefore, we have
$X \asigneq \REG^{\tau_1}[x]$ (assertion R).

In the following we show that $\REG^{\tau_1}[x] \asigneq  \REG^{\tau}[x]$.
If $\REG[x]$ was not updated between $\tau$ and $\tau_1$, then
$ \REG^{\tau_1}[x] =  \REG^{\tau}[x]$ and the claim follows.
Otherwise, if $\REG[x]$ was updated between $\tau$ and $\tau_1$,
the content of $\REG^{\tau_1}[x]$, let it be $Y$, is a result of a write
$\wrt(x,Y)$  that occurred between $\tau$ and $\tau_1$ and that was not
covering a register of $\REG$ at time $\tau$
(remember that no write is covering $\REG[x]$ at time $\tau$).
We assumed above that $\tau_2$ is the first time at which the lemma is
contradicted. Hence the write $\wrt(x,Y)$, which occurs before $\tau_2$,
complies with the requirements of the lemma. It follows that
$Y \asigneq \REG^{\tau}[x]$, and we consequently have
$\REG^{\tau_1}[x] \asigneq  \REG^{\tau}[x]$.

But it was shown above (see assertion R) that $X \asigneq \REG^{\tau_1}[x]$.
Hence, due to the transitivity of the relation $\asigneq$
(Lemma~\ref{lemma:partial-order}),  we obtain  $X \asigneq \REG^{\tau}[x]$,
a  contradiction that concludes the proof of the lemma.
\renewcommand{\toto}{lemma:tool}
\end{proofL}

%----------------------------------------------------------------------

\begin{lemma}
\label{lemma:homogeneous-snapshots-abstract}
Let $\tau$ and $\tau^\prime \geq \tau$  be two time instants.
If $\REG^{\tau^\prime}$ is $\homog(Y)$, then there exists
$X \in \REG^{\tau}$ such that $Y \asigneq X$.
\end{lemma}

\begin{proofL}
If $\REG^{\tau^\prime}=\REG^{\tau}$, the lemma holds trivially.
So let us assume in the following that
$\REG^{\tau^\prime} \neq \REG^{\tau}$ which means that a write
happens between $\tau$ and $\tau^\prime$.
If $\langle 0, \down, \false, \bot\rangle \in \REG^{\tau}$, as
every quadruplet $Y$ written in $\REG$ is such that $Y.rd \geq 1$
(line~\ref{Z-04},~\ref{Z-05}, or lines~\ref{Z-06}-\ref{Z-08}),
we have  $Y \asign \langle 0, \down, \false, \bot\rangle$.

So, let us assume that
$\langle 0, \down, \false, \bot\rangle \not\in \REG^{\tau}$
and consider the last write in $\REG$ before $\tau$.
Assume this happens at $\tau^{-} \leq \tau$ and let $p_i$
be the writing process.
Process $p_i$ has no write covering a register of $\REG$ at time $\tau^{-}$.
Consequently, at most $(n-1)$ processes\footnote{Let us notice that this
is the only
place in the proof where the consensus version of the algorithm requires more
than $(n-1)$ MWMR atomic registers.} have a write covering a register
of $\REG$ at time $\tau^{-}$.
Hence,  there exists $x \in \{1, \ldots, n\}$ such that
%$\REG^{\tau^{-}}[x]=X$ and
no write is  covering $\REG[x]$ at time $\tau^{-}$.
Let $X=\REG^{\tau^{-}}[x]=\REG^{\tau}[x]$.
If $X=Y$ then the claim of the lemma follows trivially.
So assume in the following that $X \neq Y$.
Since $\REG^{\tau^{-}}[x]=X$,
$\REG^{\tau^{\prime}}[x]=Y$ and $Y \neq X$,
there is necessarily a write $\wrt(x,Y)$ that occurred
between $\tau^-$ and $\tau^\prime$.
As this write was not  covering a register of $\REG$ at time $\tau^{-}$,
it follows (according to Lemma~\ref{lemma:tool}) that $Y \asigneq X$,
which proves the lemma.
\renewcommand{\toto}{lemma:homogeneous-snapshots-abstract}
\end{proofL}

\noindent
The following two lemmata are corollaries of
Lemma~\ref{lemma:homogeneous-snapshots-abstract}.

\begin{lemma}
\label{lemma:homogeneous-snapshots}
If $\REG^{\tau}$ is  $\homog(X)$,  $\REG^{\tau^\prime}$ is $\homog(Y)$,
and  $\tau^\prime \geq \tau$, then $Y \asigneq X$.
\end{lemma}

%------------------------------------------------------------------
\begin{lemma}
\label{lemma:after-homog-noconflict}
If $\REG^{\tau}$ is  ${H}(X)$,  $\REG^{\tau^\prime}$ is $\homog(Y)$,
$\tau^\prime \geq \tau$, $(Y.rd=X.rd)$ and $(\neg Y.\cfl)$
 then $(Y.\val=X.\val)$.
\end{lemma}

\begin{proofL}
According to Lemma~\ref{lemma:homogeneous-snapshots}, $Y \asigneq X$.
If $Y=X$ then the claim follows immediately.
So let us assume $Y \asign X$. As  $(Y.rd=X.rd)$ and $(\neg Y.\cfl)$,
the definition of $\asigneq$ implies that $Y.\val=X.\val$.
\renewcommand{\toto}{lemma:after-homog-noconflict}
\end{proofL}
%--------------------------------------------------------------------

\subsection{Exploiting homogeneous snapshots}
\label{sec:exploiting-homo-snapshots}

\begin{lemma}
\label{lemma:up-level-snapshot}
$[(X \in \REG^{\tau})~\wedge~ (X.\lvl=\up)]$
$\Rightarrow$
\big($\exists~\tau^\prime < \tau$: $REG^{\tau^\prime}$ is
$\homog(Z)$, where $Z=\langle X.rd-1, \down, \false, X.\val \rangle$\big).
\end{lemma}

\begin{proofL}
Let us first show that there is a process that writes the quadruplet
$X^\prime$ into $\REG$, with
$X^\prime= \langle X.rd, X.\lvl, \false, X.\val \rangle$.
We have two cases depending on the value of $X.\cfl$.

\begin{itemize}
\item If $X.\cfl=\false$, then let $X^\prime=X$.
Since $X.\lvl=X^\prime.\lvl=\up$,
$X$ was necessarily written into $\REG$ by some process
(let us remember that the initial value of each register of $\REG$
is $\langle 0, \down, \false, \bot\rangle$).

\item If $X.\cfl = \true$, let us consider the time $\tau_1$
at which $X$ was written for the first time into $\REG$, say by $p_i$.
Since $X.\lvl=\up$, both $\tau_1$ and $p_i$ are well defined.
This write of $X$ happens necessarily at line~\ref{Z-08}
(If it was at line~\ref{Z-04} or~\ref{Z-05}, we would have $X.\cfl = \false$).

Therefore, $X$ was computed at line~\ref{Z-06} by the function ${\sf sup}()$.
Namely we have  $X={\sf sup}(T)$,
where the set $T$ is equal to $\{view^{\tau}[1],\cdots, view^{\tau}[n],
      \langle 1,\down,\false, v_i\rangle\}$.
Observe that $X \not\in T$, otherwise $X$ would not be written for the
first time at $\tau_1$.
Let $X^\prime={\sf max}(T)$.
Since $X \not\in T$, it follows that $X \neq X^\prime$.
Due to line S6 of the function ${\sf sup}()$,
$X$ and $X^\prime$ differ only in their conflict field.
Therefore, as $X.\cfl=\true$, it follows that $X^\prime.\cfl=\false$.
Finally, as $X^\prime.\lvl=\up$ and all registers of
$\REG$ are initialized to  $\langle 0, \down, \false, \bot\rangle$,
it follows that $X^\prime$ was  necessarily
written into $\REG$ by some process.
\end{itemize}

In both cases, there exists a time at which a process writes
$X^\prime= \langle X.rd, X.\lvl, \false, X.\val \rangle$ into $\REG$.
Let us consider the first process $p_i$ that does so.
This occurs at some time $\tau_2 < \tau$. As  $X^\prime.\lvl=\up$,
this write can occur only at line~\ref{Z-04} or line~\ref{Z-08}.

We show first that this write occurs necessarily at line~\ref{Z-04}.
Assume for contradiction that the write of $X^\prime$ into $\REG$
happens at line ~\ref{Z-08}. In this case, the quadruplet $X^\prime$
was computed at line~\ref{Z-06}.
Therefore, $X^\prime = {\sf sup}(T)$ where
where the set $T$ is equal to
$\{view^{\tau_2}[1],\cdots, view^{\tau_2}[n], \langle 1,\down,\false,
 v_i\rangle\}$.
Observe that ${\sf sup}(T)$ and ${\sf max}(T)$
can differ only in their conflict field.
As ${\sf sup}(T).\cfl=X^\prime.\cfl=\false$, it follows that
$X^\prime= {\sf sup}(T)= {\sf max}(T)$.
Consequently, $X^\prime \in view^{\tau_2}$.
That is, $p_i$ is not
the first process that writes $X^\prime$ in $\REG$, contradiction.
Therefore, the write necessarily happens at line~\ref{Z-04}.

It follows then from the precondition of line~\ref{Z-04}
%(namely the snapshot value is such that  $\level=\down$)
that $view^{\tau_2}$ is
$\homog(\langle X^\prime.rd-1, \down, \false, X^\prime.\val\rangle)$.
Hence, the lemma follows.
\renewcommand{\toto}{lemma:up-level-snapshot}
\end{proofL}

%-------------------------------------------------------------
\begin{lemma}
\label{lemma:stable-snapshots}
$[(\REG^{\tau}$\text{ is }$\homog(X))$ $\wedge$
$(X.\lvl=\up)$ $\wedge$  $(\neg X.\cfl)$ $\wedge$
$(\REG^{\tau^\prime}$\text{ is }$\homog(Y))$  $\wedge$
$(Y.rd \geq X.rd)]$ $\Rightarrow$ $(Y.\val=X.\val)$.
\end{lemma}

\begin{proofL}
The proof is by induction on $Y.rd$. Let us first assume that
$Y.rd= X.rd$, for which we consider two  cases.
\begin{itemize}
\vspace{-0.2cm}
\item Case 1:  $\tau \geq \tau^\prime$.
Since $X.\cfl=\false$, it follows according to
Lemma~\ref{lemma:after-homog-noconflict} that
$Y.\val=X.\val$.

\vspace{-0.2cm}
\item Case 2: $\tau^\prime > \tau$.
According to Lemma~\ref{lemma:homogeneous-snapshots},
$Y \asigneq X$.
As $Y.rd=X.rd$, it follows that $Y.\lvl \geq X.\lvl=\up$, and consequently
$Y.\lvl=\up$.

Summarizing
we have $\REG^{\tau^\prime}$ is $\homog(Y)$, $Y.\lvl=\up$ and $Y.rd=X.rd$.
%%Assume for contradiction, the existence of $Z \in \REG^{\tau^\prime}$
%%such that $Z.\val \neq X.\val$.
%%Note that $Z.rd=X.rd$ and $Z.\lvl=X.\lvl=\up$.
According to Lemma~\ref{lemma:up-level-snapshot}, This implies
that it exists $\tau_1 < \tau$ and $\tau_1^\prime<\tau^\prime$ such hat
 $REG^{\tau_1}$ is $\homog(\langle X.rd-1, \down, \false,  X.\val\rangle)$
and $REG^{\tau_1^\prime}$ is
$\homog(\langle Y.rd-1, \down, \false, Y.\val\rangle)$.
According to Lemma~\ref{lemma:homogeneous-snapshots}, we have either
$\langle X.rd-1, \down,  \false, X.\val\rangle
  \asigneq \langle Z.rd-1, \down,  \false, Y.\val \rangle$
or $\langle Y.rd-1, \down,  \false, Y.\val\rangle
\asigneq \langle X.rd-1, \down,  \false, X.\val\rangle$.
Since by assumption $X.rd=Y.rd$, it follows
that  $X.\val=Y.\val$. The contradiction establishes the claim.
\end{itemize}

For the induction step, let assume that the lemma is true up to
$Y.rd=\rho \geq r$,  and let us prove it for $\rho+1$.
To this end, we have to show that $Y.\val=X.\val$ for every $Y$ that is
written in $\REG$ with $Y.rd=\rho+1$.
Let us assume by contradiction that $Y.\val\neq X.\val$ and
let $p_i$ be the first process that writes
$\langle \rho+1, -,-, Y.\val \rangle$ into $\REG$.
This happens at line~\ref{Z-04} or~\ref{Z-05}.
In all cases, this implies that, at this moment,
$view_j$ is $\homog(\langle\rho, -, -,Y.\val\rangle)$.
But, according to the induction assumption, this implies $Y.\val=X.\val$,
a contradiction which completes the proof of the lemma.
\renewcommand{\toto}{lemma:stable-snapshots}
\end{proofL}

%-----------------------------------------------------------------------
\subsection{Proof of the algorithm: exploiting the previous lemmas}
\label{sec:proof-algo}

\begin{lemma}
\label{lemma:agreement}
No two processes decide different values.
\end{lemma}

\begin{proofL}
Let $r$ be the smallest round in which a process decides,  $p_i$ and
$\val$ being the deciding process and the decided value, respectively.
Therefore, there is a time $\tau$ at which $view_i^\tau$ is
$\homog(\langle r, \up,\false, \val\rangle)$.
Due to  Lemma~\ref{lemma:stable-snapshots}, every homogeneous
snapshot starting from round $r$ is necessarily associated with the value
$\val$. Therefore, only this value can be decided in any round higher than $r$.
Since $r$ was assumed to be the smallest round in which a decision occurs,
the consensus agreement property follows.
\renewcommand{\toto}{lemma:agreement}
\end{proofL}
%---------------------------------------------------------------

\begin{lemma}
\label{lemma:writtenIsProposed}
For every quadruplet $X$ that is written in $\REG$,
$X.\val$ is a value proposed by some process.
\end{lemma}

\begin{proofL}
Let us assume by  contradiction that  $X.\val=v$ was not proposed by a
process, and let  $p_i$ be the first process that writes $X$ into $\REG$.
We consider  two cases according to the line at which the write occurs.
\begin{itemize}
\vspace{-0.2cm}
\item  $v$ is written  into $\REG$ at line~\ref{Z-04} or line~\ref{Z-05}.
In this case, $p_i$ obtained a view of $\REG$ in which at least some
register contains
the value $v$. According to the predicate  of these two lines,
the round number associated with $v$ is necessarily greater than $0$
 which implies
that $v$ was previously written into $\REG$ and was not there initially.
But this means that $p_i$ is not the first process
which writes $v$ into $\REG$, a contradiction.
%%\item  $v$ is written  into $\REG$ at line~\ref{Z-04} or line~\ref{Z-05}.
%%In this case, $p_i$ obtained a view of $\REG$ in which every register contains
%%the value $v$. According to the predicate  of line~\ref{Z-03},
%%$v$ is different from $\bot$, and was previously written into $\REG$
%%at least $n$ times. But this means that $p_i$ is not the first process
%%which writes $v$ into $\REG$, a contradiction.
\vspace{-0.2cm}
\item $v$ is written   into $\REG$ at line~\ref{Z-08}.
In this case,  the quadruplet $X$, where $X.\val=v$,  was returned by the
call of the function ${\sf sup}()$, namely
${\sf sup}(view[1],\cdots,view[n], \langle 1,\down,\false, v_i\rangle)$,
from which it follows that $v$ is either $v_i$ (the proposal of $p_i$)
or some value that was previously written by another process.
But, by assumption, $p_i$ is assumed to be the first process to write $v$.
Hence,  $v=v_i$, which concludes the proof of the lemma.
\end{itemize}
\vspace{-0.6cm}
\renewcommand{\toto}{lemma:writtenIsProposed}
\end{proofL}

\begin{lemma}
\label{lemma:validity}
A decided value is a proposed value.
\end{lemma}

\begin{proofL}
If a process decides a value $v$, it does it at line~\ref{Z-03}.
Hence, according to the predicate  of line~\ref{Z-03},
the round number associated with this value is greater than 0
which means that $v$ was necessarily written into $\REG$ by some process.
It then follows from  Lemma~\ref{lemma:writtenIsProposed}, that
 $v$ was proposed by a process, which  establishes the claim.
\renewcommand{\toto}{lemma:validity}
\end{proofL}
%---------------------------------------------------------------

\begin{lemma}
\label{lemma:sup-subset}
Let $T$ be a set of quadruplets.
For every $T^\prime \subseteq T:
                  {\sf sup}(T^\prime \cup \{{\sf sup}(T)\})={\sf sup}(T)$.
\end{lemma}

\begin{proofL}
Let $S= {\sf sup}(T)$. Hence $S.rd$ is the highest round number in $T$.
Moreover, $S$ is greater than, or equal to, any quadruplet in $T$.
Hence, ${\sf max}(T^\prime \cup \{S\})=S$.
Therefore, combined with the the definition of ${\sf sup}()$,  we have:
%\begin{align*}
%{\sf sup}(T^\prime \cup \{S\})= &\langle S.rd, \conflict, S.\val\rangle
%                                &\text{ if } \conflictf(T^\prime \cup \{S\})\\
%= &\langle S.rd, S.\lvl, S.\val\rangle &\text{ otherwise.}
%\end{align*}
%\begin{align*}
${\sf sup}(T^\prime \cup \{S\})= \langle S.rd, S.\lvl,
        \conflictf(T^\prime \cup \{S\}) , S.\val\rangle$.
%\end{align*}
Thus, in order to prove that ${\sf sup}(T^\prime \cup \{S\})=S$, we need to show
that $\conflictf(T^\prime \cup \{S\}) =S.\cfl$.
There are two cases depending on the value of $S.\cfl$.
\begin{itemize}
\vspace{-0.2cm}
\item  $S.\cfl= \true$. \\
In this case,   $\conflictOne(\{S\})=\true$.
But $S.rd$ is the highest round number in $T$ from which it follows that
$S.rd$ is also the highest in $T^\prime \cup \{S\}$.
Therefore,  $\conflictOne(\{S\})=\true$ implies that
$\conflictOne(T^\prime \cup \{S\})=\true$.
\vspace{-0.2cm}
\item  $S.\cfl=\false$.\\
Since $S={\sf sup}(T)$, it follows that $\conflictf(T) = \false$.
Consequently,
both $\conflictOne(T)$ and $\conflictTwo(T)$ are $\false$.
Moreover, as $S.\cfl=\false$, it follows that $\conflictOne(\{S\})=\false$.
Therefore $\conflictOne(T \cup \{S\})=\false$.
But, as  $T^\prime \subseteq T$,
this yields $\conflictOne(T^\prime \cup \{S\})=\false$.

On another side, it follows from $\conflictTwo(T)=\false$ that $|vals(T)|=1$.
As $S={\sf sup}(T)$, we have  $S.\val \in vals(T)$.
Therefore $|vals(T \cup \{S\})|=1$.
Since $T^\prime \subseteq T$, it follows that
$|vals(T^\prime \cup \{S\})|=1$ which implies
$\conflictTwo(T^\prime \cup \{S\})=\false$.

As both $\conflictOne(T^\prime \cup \{S\})$ and
$\conflictTwo(T^\prime \cup \{S\})$ are false, it follows that
$\conflictf(T^\prime \cup \{S\})=\false$.
\end{itemize}
From the case analysis we conclude that
$\conflictf(T^\prime \cup \{S\}) = S.\cfl$.
\renewcommand{\toto}{lemma:sup-subset}
\end{proofL}

\begin{lemma}
\label{lemma:OB-termination}
If there is a time after which a process executes solo, it decides a value.
\end{lemma}

\begin{proofL}
Assume that $p_i$ eventually runs solo, we need to show that $p_i$ decides.
There exists a time $\tau$, after which no other process than $p_i$ writes
into $\REG$.
Let $\tau^\prime \geq \tau$ be the first time at which $p_i$ takes a
snapshot after $\tau$. This snapshot is well defined, as $p_i$ runs solo
after $\tau$ and the  implementation of atomic snapshot is obstruction-free.
Let $S={\sf sup}(view_i^{\tau^\prime}[1],\cdots,view_i^{\tau^\prime}[n],
\langle 1,\down, \false, v_i\rangle)$.

Let us first show that there is a time after $\tau$ at which
$\REG$ is $\homog(S)$.

\begin{itemize}
\vspace{-0.2cm}
\item
If $\REG^{\tau^\prime}$ is $\homog(S)$, we are done.
\vspace{-0.2cm}
\item
If $\REG^{\tau^\prime}$ is not $\homog(S)$, $p_i$ executes line~\ref{Z-06} and
computes $S$. Then it writes $S$ in an entry of $\REG$
(containing a value different from $S$),
and re-enters the loop.   If $\REG$ is then $\homog(S)$, we are done.
Otherwise, $p_i$ executes again line~\ref{Z-06} and, due to
Lemma~\ref{lemma:sup-subset},  the quadruplet computed by the function
${\sf sup}()$ is equal to $S$.
It follows that after a finite number of iterations of the loop,
$\REG$ is  $\homog(S)$.
\end{itemize}
When $\REG$ is  $\homog(S)$, we have the following.
\begin{itemize}
\vspace{-0.2cm}
\item
If $S=\langle -, \up,  \false, - \rangle$, $p_i$ decides in line~\ref{Z-03}.
\vspace{-0.2cm}
\item If $S=\langle r, \down, \false, \val \rangle$, then $p_i$ writes
$Y=\langle r+1, \up,  \false, \val \rangle$ in
line~\ref{Z-04}. Using the same argument as above, there is a time at which
$\REG$ becomes $\homog(Y)$, and the previous case holds.
\vspace{-0.2cm}
\item If $S=\langle r, -, \true, \val\rangle$, then
$p_i$ writes $Y=\langle r+1, \down, \false, \val \rangle$ in line~\ref{Z-05}.
% where $\val$ is any value in $\REG$.
Then $p_i$ keeps writing $Y$ in the following iterations until $\REG$ becomes
$\homog(Y)$,  and the previous case holds.
\end{itemize}
\vspace{-0.1cm}
Hence, in all cases $p_i$ eventually decides.
\renewcommand{\toto}{lemma:OB-termination}
\end{proofL}
%---------------------------------------------------------------

\begin{lemma}
\label{lemma:SV-termination}
If a single value is proposed, all correct processes decide.
\end{lemma}

\begin{proofL}
Let us assume that all processes propose the same value $v$.
It follows that  all the processes keep writing
$X= \langle 1, \down, \false, v \rangle$ until $\REG$ becomes $\homog(X)$.
Then, once every register of $\REG$ has been updated at least once,
the processes start writing $Y= \langle 2, \up, \false, v \rangle$ until
$\REG$ becomes $\homog(Y)$ and $v$. When this occurs, $v$ is  decided.
\renewcommand{\toto}{lemma:SV-termination}
\end{proofL}

%---------------------------------------------------------------
\begin{theorem}
\label{theo:OB-consensus}
The algorithm described in Figure~{\em\ref{fig:algorithm-OB-consensus}}
solves the obstruction-free consensus problem
(as defined in Section~{\em\ref{sec:OB-consensus-definition}}).
\end{theorem}

\begin{proofT}
The proof follows directly from the
Lemma~\ref{lemma:agreement} (Agreement),
Lemma~\ref{lemma:validity} (Validity),
Lemma~\ref{lemma:OB-termination} (OB-Termination),
and Lemma~\ref{lemma:SV-termination} (SV-Termination).
\renewcommand{\toto}{theo:OB-consensus}
\end{proofT}

%\vspace{-0.2cm}
%======================================================================
\section{From Consensus to $(n,k)$-Set Agreement}
\label{sec:k-set-agreement}
\vspace{-0.3cm}
\paragraph{The algorithm}
The  obstruction-free $(n,k)$-set agreement algorithm is the same as the one
 of Figure~\ref{fig:algorithm-OB-consensus}, except that now there are
only $m=n-k+1$ MWMR atomic registers instead of $m=n$. Hence $\REG$ is now
$\REG[1..(n-k+1)]$.

\paragraph{Its correctness}
The arguments for the validity and liveness properties are the same  as the
ones of the  consensus algorithm since they do not depend on the size of the
memory $\REG$.

As far as the $k$-set agreement  property is concerned (no more than
 $k$ different values can be  decided),  we have to
show that $(n-k+1)$  registers are sufficient.  To this end, let us
consider the $(k-1)$ first decided values, where
the notion ``first'' is defined with respect to the linearization time
of the snapshot invocation (line~\ref{Z-02}) that immediately precedes the
invocation of the corresponding deciding statement
(${\sf return}()$ at line~\ref{Z-04}). Let $\tau$ be the time  just after
the linearization of these   $(k-1)$ ``deciding'' snapshots.
Starting from $\tau$, at most $(n-(k-1))=(n-k+1)$ processes access the array
$\REG$, which is made up of exactly $(n-k+1)$ registers.
Hence, after $\tau$, these $(n-k+1)$ processes execute the consensus
algorithm of Figure~\ref{fig:algorithm-OB-consensus}, where
$(n-k+1)$ replaces $n$, and consequently at  most one new value is decided.
Therefore, at most $k$ values are decided by the $n$ processes.

\vspace{-0.1cm}
%========================================================================
\section{From One-shot to Repeated $(n,k)$-Set Agreement}
\label{sec:repeated}
\vspace{-0.1cm}
\subsection{The repeated  $(n,k)$-set agreement problem}
In the repeated  $(n,k)$-set agreement problem, the processes executes a
sequence of  $(n,k)$-set agreement instances.
Hence, a process $p_i$ invokes sequentially
the  operation ${\sf propose}(1,v_i)$, then ${\sf propose}(2,v_i)$, etc.,
where $sn_i=1,~2,~...$  is the sequence number of its current instance,
and $v_i$ the value it proposes to this instance.

It would be possible to associate a specific  instance of the base
algorithm  described in Figure~\ref{fig:algorithm-OB-consensus}
with each sequence number, but this would require $(n-k+1)$ atomic read/write
registers per instance. The next section that, it is possible to
solve the repeated problem with only $(n-k+1)$ atomic registers.
According to the complexity results of~\cite{DFKR15}, it follows that
this algorithm is optimal in the number of atomic registers, which
consequently closes the lower/upper bounds discussion associated with
repeated $(n,k)$-set agreement.

%---------------------------------------------------------------------------
\subsection{Adapting the algorithm}

\paragraph{From quadruplets to sixuplets}
Instead of a quadruplet, an atomic read/write register is now a sixuplet
$X=\langle sn, rd, \lvl, \cfl, \val, dcd \rangle$. The
four fields $X.rd$, $X.\lvl$, $X.\cfl$, $X.\val$ are the same as before.
The new field $X.sn$ contains a sequence number, while the new field
$X.dcd$ is an initially empty list.  From a notational point of view,
the $j$th element of this list is denoted $X.dcd[j]$; it contains a value
decided by the $j$th instance of the repeated $(n,k)$-set agreement.

The total order on sixuplets ``$>$'' is the classical lexicographical order
defined on its first five fields while the relation ``$\asign$'' is now
defined as follows:
$$\arel{X}{Y}~ \stackrel{\mathit{def}}{=}~
(X >Y)\wedge
[{\color{red}(X.sn > Y.sn)} \vee (X.rd > Y.rd)\vee (X.\cfl)
                              \vee(\neg Y.\cfl \wedge X.\val=Y.\val)].$$

\vspace{-0.2cm}
\paragraph{Local variables}
Each process $p_i$ has now to manage two local variables whose scope is
the whole  repeated  $(n,k)$-set agreement problem.
\begin{itemize}
\vspace{-0.2cm}
\item The variable $sn_i$, initialized to $0$,  is used by $p_i$
to generate its sequence numbers. It is assumed that $p_i$ increases
 $sn_i$ before invoking ${\sf propose}(sn_i,v_i)$.
\vspace{-0.2cm}
\item  The local list $dcd_i$ is used by $p_i$ to store the value it has
decided during the previous instances of the $(n,k)$-set agreement.
Hence,  $dcd_i[j]$ contains the value decided by $p_i$ during the
$j$th instance.
\end{itemize}

\paragraph{The algorithm}
The algorithm executed by a process $p_i$
is described in Figure~\ref{fig:algorithm-OB-consensus-repeated}.
The parts which are new with respect to the base algorithm of
Figure~\ref{fig:algorithm-OB-consensus} are in red.

%========================================================================
\begin{figure}[ht]
\centering{
\fbox{
\begin{minipage}[t]{150mm}
\footnotesize
\renewcommand{\baselinestretch}{2.5}
\resetline
\begin{tabbing}
aaaaa\=a\=aa\=aaa\=aaaaa\=aaaaa\=aaaaaaaaaaaaaa\=aaaaa\=\kill %~\\

{\bf operation} ${\sf propose}(\sni, v_i)$ {\bf is} \\

\line{R-01} \> {\bf repeat forever} \\

\line{R-02} \>\> $view \leftarrow \REG.{\sf snapshot}();$\\

\line{R-03} \> \>
   {\bf case} \= $(\forall x: view[x]=\langle \sni, r,\up,\false,\val, -\rangle
                          ~~~~~\text{where }r > 0)$
   ~\={\bf then} {\color{red}$\dcd_i[\sn_i] \gets \val$;}
                 ${\sf return}(\val$)\\

\line{R-04} \>\>\>
     $(\forall x: view[x]=\langle \sni, r,\down,\false,\val, -\rangle
            ~\text{where }r > 0)$ \>
 {\bf then}
     $\REG.{\sf write}(1, \langle \sni, r+1,\up,\false, \val, \dcdi\rangle)$\\

\line{R-05} \>\>\>
  $(\forall x: view[x]=\langle \sni, r,\level,\true,\val, -\rangle
               ~~\text{where }r > 0)$  \>

% {\bf then} \=
%    {\bf let} $w$ {\bf be} any value in $view$ different from $\bot$;\\

%\line{R-06} \>\>\>\>\>
    $\REG.{\sf write}(1, \langle \sni, r+1,\down,\false, \val, \dcdi\rangle)$\\

\line{R-06} \>\>\> {\bf otherwise} \=
{\bf let}
 $ \langle {\color{red}inst}, r,\level,\conflict, \val, \dec \rangle
   \leftarrow
  {\sf sup}(view[1],\cdots,view[n], \langle \sni, 1,
                          \down, \false, v_i, \dcdi\rangle)$;\\

\line{R-07} \>\>\>\>
{\color{red}{\bf if} $(inst > \sni)$
     {\bf then} $\dcdi[\sni] \gets \dec[\sni]$;
     ${\sf return} ~\dcdi[\sni]$  {\bf end if}} \\

\line{R-08}  \>\> \>\>
    $x \gets$
    smallest index such that $view[x]=~{\sf min}(view[1],\cdots,view[n])$;\\

\line{R-09}  \>\> \>\>
$\REG.{\sf write}(x,
  \langle {\color{red}inst}, r,\level, \conflict,\val, \dec\rangle)$\\
% $\REG.{\sf write}(x, \langle {\color{red}inst} r,\level,\conflict, \val, {\color{red}\dec}\rangle)$\\

\line{R-10} \>\> {\bf end case}\\

\line{R-11} {\bf end repeat}.

\end{tabbing}
\normalsize
\end{minipage}
}
\caption{Repeated obstruction-free Consensus}
\label{fig:algorithm-OB-consensus-repeated}
}
\end{figure}
%=======================================================================

\begin{itemize}
\vspace{-0.2cm}
\item Line~\ref{R-03}.
When all entries of a view obtained by $p_i$ contain only  sixuplets
whose the first five fields are equal, $p_i$ decide the value $\val$.
But before returning $\val$, $p_i$ writes it in $dcd_i[sn_i]$.
Hence, when  $p_i$ will execute the next $(n,k)$-set agreement instance
(whose occurrence number will be $sn_i+1$),
it will be able to help processes, whose current sequence  number $sn'$
are smaller than $sn_i$,  decide a value returned by the instance  $sn'$
of the repeated  $(n,k)$-set agreement.
\vspace{-0.2cm}
\item Line~\ref{R-04}.
In this case, $p_i$  obtains a view whose five first entries are equal to
$\langle \sni, r,\down,\false,\val\rangle$.  It then writes
in $\REG[1]$ the sixuplet $\langle sn_i, r,\down,\false,\val, dcd_i\rangle$.
Let us notice that the write of $dcd_i$ is to help other processes
decides in $(n,k)$-set agreement instances whose sequence number is
smaller than $sn_i$.
\vspace{-0.2cm}
\item Line~\ref{R-05}.
This case is similar to the previous one.
\vspace{-0.2cm}
\item Lines~\ref{R-06}-\ref{R-10}.
In this case, $p_i$ computes the supremum of the snapshot value $view$
obtained at line~\ref{R-03} plus the qsixuplet
$\langle sn_i, 1,\down,\false,\val, dcd_i\rangle$. There are two cases.
\begin{itemize}
\vspace{-0.2cm}
\item
If the sequence number of this supremum $inst$ is greater than $sn_i$
(line~\ref{R-07}), $p_i$ can benefit from the list of values already
decided in   $(n,k)$-set agreement instances whose sequence number
is smaller than $inst$. This help is  obtained from $dec[sn_i]$.
Consequently, similarly to line~\ref{R-03}, $p_i$ writes this value in
$dcd_i[sn_i]$ and decides it.
\item
If $inst=sn_i$, $p_i$ executes as in the  base algorithm
(lines~\ref{R-08}-\ref{R-09}).
\end{itemize}
\end{itemize}
\vspace{-0.2cm}
Hence, solving repeated $(n,k)$-set agreement in an anonymous system does not
require more atomic read/write registers than the base non-repeated version.
The only additional cost lies in the size of the atomic registers
which contain two supplementary unbounded fields. As already indicated,
it follows from the lower bound established in~\cite{DFKR15} that this
algorithm is optimal with respect to the number of underlying atomic registers.

%======================================================================

\section{From Obstruction-Freedom to $x$-Obstruction-Freedom}
\label{sec:x-obstruction-freedom}

This section extends the base algorithm to obtain an algorithm
that solves the $x$-obstruction-free $(n,k)$-set agreement problem.
Let $x\leq k$ (\footnote{This assumption is a necessary requirement to solve
$(n,k)$-set agreement in a read/write system. It follows from the impossibility
result stating that $(n,k)$-set agreement cannot be wait-free solved for
$n > k$, when any number of processes may crash~\cite{BG93,HS99,SZ00}.}).

\paragraph{One-shot $x$-obstruction-freedom}
This progress condition, introduced in~\cite{T09,T09-Opodis},  is a natural
generalization of obstruction-freedom, which corresponds to the case $x=1$.

$x$-Obstruction-freedom guarantees that,  for every set of processes
$P$, $|P|\leq x$,  every correct process in $P$ returns from
its operation invocation  if no process outside $P$  takes steps for ``long
enough''.  It is easy to see that  $x$-obstruction-freedom and wait-freedom
are equivalent  in any $n$-process system where $x \geq n$.
Differently, when $x<n$, $x$-obstruction-freedom depends on the concurrency
pattern while wait-freedom does not.

\paragraph{$x$-Obstruction-free $(n,k)$-set agreement: OB-Termination}
When considering  $x$-obstruction-freedom, the
Validity, Agreement and SV-Termination properties defining
obstruction-free $(n,k)$-set agreement  are the same as the
ones stated in Section~\ref{sec:OB-k-set-agreement-definition}.
The only property that must be adapted is OB-Termination, which becomes:
\begin{itemize}
\vspace{-0.1cm}
\item $x$-OB-termination.
If there is a time after which at most $x$ correct processes  execute
concurrently, each of these processes eventually decides a value.
\end{itemize}

\paragraph{The shared memory $\REG$}
To cope with the $x$-concurrency allowed by obstruction-freedom,
the array $\REG$ is such that it has now $m=n-k+x$ entries
(i.e.,  $m=n-k+1)$ entries for the base obstruction-freedom).
This increase in the size of the array is due to the fact that
the algorithm is required to terminate in more scenarios than
simple obstruction-freedom.

\paragraph{Content of a quadruplet}
In the base algorithm, the four fields of a quadruplet $X$  are
a round number $X.rd$, a level $X.\lvl$, a conflict value $X.cf\ell$,
and a value  $X.val$.
Coping with $x$-concurrency requires to replace the last field, which was
made up of a single $X.val$,  by a set of values denoted $X.valset$.

%========================================================================
\begin{figure}[ht]
\centering{
\fbox{
\begin{minipage}[t]{150mm}
\footnotesize
\renewcommand{\baselinestretch}{2.5}
\resetline
\begin{tabbing}
aaaaaaa\=aa\=aaa\=aaa\=aaaaa\=aaaaa\=aaaaaaaaaaaaaa\=aaaaa\=\kill %~\\

{\bf function}  ${\sf sup}(T)$ {\bf is}
 ~~\%  $S$ is a set of quadruplets,  the last field
           of each of them is now a set of values \% \\

(S1') \> {\bf let}  $\langle r,\level,\cfl, valset \rangle$
            {\bf be}  ${\sf max}(T)$;
~~~~~~~~~~~~~~~~~\% lexicographical order \% \\

(S2') \> {\bf let} $vals(T)$  ~{\bf be}
       $\{ v ~|~ \langle r,-,valset\rangle \in T ~\wedge~ v\in valset \}$; \\

%%(N)  \>  {\bf let} $vals'(T)$ {\bf be}
%%               the set of the (at most) $x$ greatest values in $vals(T)$;\\

(S3) \> {\bf let} $\conflictOne(T)$  {\bf be}
             $\exists ~\langle r, -, \conflict, -\rangle \in T$;
             ~~~~~~\% conflict inherited \% \\

(S4') \> {\bf let} $\conflictTwo(T)$  {\bf be} $|vals(T)| >x$;
             ~~~~~~~~~~~~~~~~~~~~~\% conflict discovered \%  \\

(S5) \> {\bf let}
$\conflictf(T)$ {\bf be} $\conflictOne(T) \vee \conflictTwo(T)$;\\

(N) \> {\bf if} $\conflictf(T)$ {\bf then} $vals'(T) \gets valset$ {\bf else}
               $vals'(T) \gets$ the set of the (at most) $x$ greatest values in $vals(T)$ {\bf end if};\\

(S6') %\> {\bf if} $\conflictf(T)$
     %            {\bf then} $\level \gets \conflict$ {\bf end if};\\

%(S7')
\> ${\sf return}
             \big(\langle r,\level,\conflictf(T),  vals'(T)\rangle\big)$.

\end{tabbing}
\normalsize
\end{minipage}
}
\caption{Function ${\sf sup}()$ suited to $x$-obstruction-freedom}
\label{fig:algorithm-sup-m-OB}
}
\end{figure}
%======================================================================

\paragraph{The modified function ${\sf sup}()$}
Coping with $x$-concurrency requires to also adapt the function ${\sf sup}()$.
This function ${\sf sup}()$ is a simple extension
of the  base version described in Figure~\ref{fig:algorithm-sup},
that allows to consider a set of values instead of a single value.
 It is described in Figure~\ref{fig:algorithm-sup-m-OB}.
The lines that are modified (with respect to the base function ${\sf sup}()$)
are followed by a ``prime'', and  a new line (marked N) is added.
More precisely, the modifications are the following.
\begin{itemize}
\vspace{-0.1cm}
\item  Line S1'.
The last field of a quadruplet is now a set of values, denoted $valset$.
As far as the  lexicographical ordering is concerned,
the sets $valset$ are ordered as follows. They  are
ordered by size, and sets of the same size are
ordered from their greatest to their smallest element.
\vspace{-0.1cm}
\item  Line S2'.
The set $vals(T)$ is now the union of all the $valset$ associated
with the greatest round number appearing in  $T$.
\vspace{-0.1cm}
\item  Lines S3 and  S5:  not modified.
\vspace{-0.1cm}
\item Line S4'.
$\conflictTwo(T)$ is modified to take into account $x$-concurrency.
A conflict is now discovered when more than $x$ (instead of $1$)
values are associated with the round number of the maximal element of $T$.
\vspace{-0.2cm}
\item New line N.
The set $vals'(T)$ is equal to $valset$ if $\conflict(T)=\true$.
Otherwise, it contains the (at most) $x$ greatest values of $vals(T)$.
\vspace{-0.1cm}
\item Line S6'.
The quadruplet returned by ${\sf sup}(T)$
differs from the one of Figure~\ref{fig:algorithm-OB-consensus}
in its last field which is now the set  $vals'(T)$.
\end{itemize}

It is easy to see that, when the last field of the quadruplets is reduced to
singleton, and $x=1$, this extended version boils down to the one described in
Figure~\ref{fig:algorithm-OB-consensus}.

%========================================================================
\begin{figure}[ht]
\centering{
\fbox{
\begin{minipage}[t]{150mm}
\footnotesize
\renewcommand{\baselinestretch}{2.5}
\resetline
\begin{tabbing}
aaaaa\=aa\=aaa\=aaa\=aaaaa\=aaaaa\=aaaaaaaaaaaaaa\=aaaaa\=\kill %~\\

{\bf operation} ${\sf propose}(v_i)$ {\bf is} \\

\line{W-01} \> $Q \gets \langle 1,\down, \false, \{v_i\}\rangle$; \\

\line{W-02} \> {\bf repeat forever} \\

\line{W-03} \>\> $view \leftarrow \REG.{\sf snapshot}();$\\

\line{W-04} \> \>
   {\bf case} \= $(\forall x: view[x]= Q = \langle r,\up,\false,valset\rangle
                          ~~~~~~~\text{where }r > 0)$
                          ~\={\bf then} ${\sf return}$ any value in $valset$\\

\line{W-05} \>\>\>$(\forall x: view[x]= Q = \langle r,\down,\false, valset \rangle
            ~~~\text{where }r > 0)$ \>
 {\bf then} $Q \gets \langle r+1,\up,\false, valset\rangle; \REG.{\sf write}(1, Q)$\\

\line{W-06} \>\>\> $(\forall x: view[x]= Q = \langle r,\level,\true, valset\rangle
               ~~~~\text{where }r > 0)$  \> {\bf then} \=
 {\bf let} $v$ be any value in $valset$; \\
\line{W-07}  \>\>\>\>\>  $Q \gets \langle r+1,\down,\false,  \{v\}\rangle$; $\REG.{\sf write}(1, Q)$;\\

%\line{W-06} \>\>\>\>\>
%    $\REG.{\sf write}(1, \langle r+1,\down,\false,  w\rangle)$\\

\line{W-08} \>\>\> {\bf otherwise} \=
{\bf let}   $Q \leftarrow
     {\sf sup}(view[1],\cdots,view[n], Q)$;\\

\line{W-09}  \>\> \>\>
    $x \gets$
    smallest index such that $view[x] \neq Q $;\\

\line{W-10}  \>\> \>\>
$\REG.{\sf write}(x, Q)$\\

\line{W-11} \>\> {\bf end case}\\

\line{W-12} {\bf end repeat}.

\end{tabbing}
\normalsize
\end{minipage}
}
\caption{Anonymous x-obstruction-free Consensus}
\label{fig:algorithm-XOB-consensus}
}
\end{figure}
%=======================================================================
\paragraph{$x$-Obstruction-free $(n,k)$-set agreement: algorithm}
An algorithm extending the base obstruction-free algorithm
of Figure~\ref{fig:algorithm-OB-consensus} to an
$x$-obstruction-free $(n,k)$-set agreement algorithm is described in
Figure~\ref{fig:algorithm-OB-consensus}.
(Let us remember that, as the underlying snapshot algorithm is
non-blocking~\cite{GR07}, it ensures that --whatever the concurrency pattern--
at least one snapshot invocation always terminates.)
This algorithm solving the $x$-obstruction-free
$(n,k)$-set  agreement problem is obtained as follows, where
(as already indicated) the array $\REG$ is composed of  $m=n-k+x$
atomic read/write registers.

\begin{itemize}
\vspace{-0.2cm}
\item
The relation  ``$\asign$'' introduced in Section~\ref{sec:proof-definitions}
is extended to take into account the fact that the last field of a
quadruplet is now a non-empty set of values. It becomes:
\begin{quote}
$\arel{X}{Y}~ \stackrel{\mathit{def}}{=}~
(X >Y)\wedge  [(X.rd > Y.rd)\vee (X.\cfl)
                 \vee(\neg Y.\cfl \wedge X.valset \supseteq Y.valset)].$
\end{quote}
\vspace{-0.2cm}
\item
Each process $p_i$ maintains a local quadruplet denoted $Q$,
containing the last quadruplet it has computed. Initially, $Q$ is equal
to  $\langle 1,\down, \false, \{v_i\}\rangle$ (line~\ref{W-01})\footnote{Let
us notice that, the algorithm has no longer the {\it memoryless} property
of the base algorithm.}.

This quadruplet allows its owner $p_i$ to have an order on the
all the quadruplets it champions during the execution of ${\sf propose}(v_i)$.
Hence, if $p_i$ champions $Q$ at time $\tau$, and  champions
$Q'$ at time $\tau'\geq\tau$, we have  $Q' \asigneq   Q$. This is to ensure the
$x$-OB-termination property.

The meaning of the three predicates at lines~\ref{W-04}-\ref{W-06},
is the following. All entries of $view$ are the same and are equal to $Q$,
where the content of $Q$ is either $\langle r,\up,\false,valset\rangle$,
or $\langle r,down,\false,valset\rangle$,
or $\langle r,\level,\true,valset\rangle$.
Hence, according to the terminology of the proof of the base algorithm,
introduced in Section~\ref{sec:proof-definitions}, $view$ is homogeneous, i.e.,
$view$ is ${\cal H}(Q)$  where $Q$ obeys some predefined pattern.
\vspace{-0.2cm}
\item
Lemma~\ref{lemma:stable-snapshots} needs to be re-formulated to take into
account the set field of each quadruplet. It becomes:
\begin{quote}
$[(\REG^{\tau}$\text{ is }$\homog(X))$ $\wedge$
$(X.\lvl=\up)$ $\wedge$  $(\neg X.\cfl)$ $\wedge$
$(\REG^{\tau^\prime}$\text{ is }$\homog(Y))$  $\wedge$
$(Y.rd \geq X.rd)]$
$\Rightarrow$ $(Y.valset \supseteq X.valset \vee X.valset \supseteq Y.valset)$.
\end{quote}
The lemma is true if the number of participating processes does not exceed
the number of available
registers in $\REG$.
\vspace{-0.2cm}
\item
As far the $k$-set agreement  property (no more than
$k$ different values can be  decided),  we have to
show that $(n-k+x)$  registers are sufficient.  The reasoning is similar
to one done at the end of Section~\ref{sec:k-set-agreement}.
More precisely,  let us consider the $(k-x)$ first decided values, where
the notion ``first'' is defined with respect to the linearization time
of the snapshot invocation (line~\ref{Z-02}) that immediately precedes the
invocation of the corresponding deciding statement
(${\sf return}()$ at line~\ref{Z-04}). Let $\tau$ be the time  just after
the linearization of these   $(k-x)$ ``deciding'' snapshots.
Starting from $\tau$, at most $(n-(k-x))=(n-k+x)$ processes access the array
$\REG$, which is made up of exactly $(n-k+x)$ registers.
Consider the $(k-x+1)$-th deciding snapshot, let it be at $\tau^\prime > \tau$.
According to the precondition of line~\ref{W-03},
$\REG^{\tau^\prime}$ is $\homog(X)$ for some $X$ with $X.\lvl=\up$
and $X.\cfl=\false$. Observe that $|X.valset| \leq x$.

According to the new statement of Lemma~\ref{lemma:stable-snapshots},
since starting from $\tau$ the number of participating
processes is always less than the number of registers, then all deciding
snapshots after $\tau^\prime$ are associated with a set of values that is
either a subset or a superset of $X.valset$.
Hence, at most $x$ values can be decided starting from $\tau^\prime$.
\vspace{-0.2cm}
\item As far as $x$-OB-termination is concerned, the key is line~\ref{W-07}.
When a process $p_i$ detects a conflict ($Q.\cfl=\true$, at line~\ref{W-06}),
it starts a new round with a set which is a singleton.
Hence, if there is a finite time after which no more than $x$ processes
are executing, there is a finite round from which at most $x$ values
survive and appear in the next round. From that round, no new conflict can
be discovered, and eventually the (at most) $x$ running processes
obtain snapshots entailing decision.
\end{itemize}

%========================================================================
\section{Conclusion}
\label{sec:conclusion}

This paper presented first a base
a one-shot obstruction-free $(n,k)$-set agreement algorithm
for a system made up of $n$ asynchronous and anonymous processes,
which communicate through atomic read/write registers.
This algorithm requires only $(n-k+1)$ such registers.
From this cost point of view, it is the best algorithm known
so far (the best previously known algorithm requires $2(n-k)+1$ atomic
read/write registers). Hence,  this algorithm answers the challenge
posed in~\cite{DFGR13}, and establishes a new upper bound
of  $(n-k+1)$ on the number of registers to solve the one-shot
obstruction-free $(n,k)$-set agreement problem. This upper bound improves
the ones stated in~\cite{DFKR15} for anonymous and non-anonymous systems.

A simple extension of the previous algorithm has then been presented,
that solves the repeated $(n,k)$-set agreement problem.
While the lower bound of $(n-k+1)$ atomic registers was established
in~\cite{DFKR15} for this problem, the proposed algorithm shows that
the upper bound is also equal $(n-k+1)$, and consequently
the proposed algorithm is optimal.
The paper has also generalized the base one-shot algorithm to
solve the $(n,k)$-set agreement problem in the context of
$x$-obstruction-freedom. The corresponding algorithm reduces
to $(n-k+x)$ the upper bound on the number of atomic read/write registers.

To attain these  goals the algorithms,
which have  been presented in an incremental way, rely on a simple
round-based structure. Moreover, the base  one-shot algorithm
does not require persistent local  variables, and, in addition to a
proposed value, an atomic register contains only two bits and a round number.
The algorithm solving the repeated $(n,k)$-set agreement problem requires
that each atomic register  includes two more integers.

Let us call ``MWMR-$nb$'' of a problem $P$, the minimal number of
MWMR atomic registers needed to solve $P$ in an asynchronous system of $n$
processes. The paper has shown that $(n-k+1)$ is the MWMR-$nb$ of
repeated obstruction-free $(n,k)$-set agreement.
We conjecture that  $(n-k+1)$ is also the MWMR-$nb$ of one-shot
obstruction-free $(n,k)$-set agreement, and more generally that
$(n-k+x)$ is the MWMR-$nb$ of one-shot $x$-obstruction-free $(n,k)$-set
agreement, when $1\leq x\leq k<n$.

%========================================================================
%\newpage
%\setcounter{page}{1}
%\pagenumbering{roman}

\section*{Acknowledgments}
This work has been partially supported by the French ANR project DISPLEXITY
devoted to computability and complexity in distributed computing, and the
Franco-German ANR project DISCMAT devoted to connections
between mathematics and distributed computing.

%========================================================================

%=======================================================================
\appendix

%=======================================================================

\section{Non-blocking snapshot object}
\label{annex:OB-free-snapshot}
This appendix presents a non-blocking (hence obstruction-free)  snapshot object
which uses no additional atomic register. The idea that underlies
this  algorithm, which is  due to Guerraoui and Ruppert~\cite{GR07}, is simple.
The algorithm, described in Figure~\ref{fig:algorithm-write-snapshot},
considers that the $n$ anonymous processes share $m$ underlying MWMR atomic
registers.

%========================================================================
\begin{figure}[ht]
\centering{
\fbox{
\begin{minipage}[t]{150mm}
\footnotesize
\renewcommand{\baselinestretch}{2.5}
\resetline
\begin{tabbing}
aaaaa\=aa\=aaa\=aaa\=aaaaa\=aaaaa\=aaaaaaaaaaaaaa\=aaaaa\=\kill %~\\

{\bf Shared variables} \\
\>  $\SM[1..m]$: array of $n$ multivalued MWMR atomic registers,
  initially  $[\langle -,\bot \rangle,\cdots,\langle -,\bot \rangle]$; \\
\>   $\SM[x]=\langle \SM[x].ts,\SM[x].value \rangle$;
     only $\SM[i].value$ can be made visible outside.  \\~\\
{\bf Permanent local variable}: each process $p_i$manages a counter $ts_i$,
                      initialized to $0$. \\~\\

{\bf operation} ${\sf write}(x,v)$ {\bf is} ~~\% issued by $p_i$~~\%\\

\line{WS-01} $\SM[x] \leftarrow \langle ts_i,v \rangle$;
             $ts_i  \leftarrow ts_i+1$;
             ${\sf return}()$.\\~\\

{\bf operation} ${\sf snapshot}()$ {\bf is} \\

\line{WS-02} \>$count \leftarrow 1$;
 {\bf for each} $x\in \{1,\ldots,m\}$  {\bf do}
               $sm1[x] \leftarrow \SM[x]$  {\bf end for};\\

\line{WS-03} \> {\bf repeat forever} \\

\line{WS-04} \> \> {\bf for each} $y\in \{1,\ldots,m\}$  {\bf do}
               $sm2[y] \leftarrow \SM[y]$  {\bf end for};\\

\line{WS-05} \> \>
   {\bf if} \= $(\forall~x\in\{1,\cdots,m\}:~ sm1[x]=sm2[x])$  \\

\line{WS-06} \> \> \> {\bf then} \= $count \leftarrow count +1$;\\

\line{WS-07} \> \> \> \> {\bf if} $(count=m(n-1)+2)$
                     {\bf then} ${\sf return}(sm1[1..m].value)$ {\bf end if}\\

\line{WS-08} \> \> \> {\bf else}  \> $count \leftarrow 1$\\

\line{WS-09} \> \> {\bf end if};\\

\line{WS-10} \> \>     $sm1[1..m] \leftarrow sm2[1..m]$ \\

\line{WS-11} {\bf end repeat}.

\end{tabbing}
\normalsize
\end{minipage}
}
\caption{Obstruction-free snapshot object  \cite{GR07}}
\label{fig:algorithm-write-snapshot}
}
\end{figure}
%=======================================================================

Each process $p_i$ manages an integer local variable $ts_i$, that it
uses to associate a sequence number to its successive write operations into
any atomic  register $\SM[x]$ (line~\ref{fig:algorithm-write-snapshot}).

When a process  invokes ${\sf snapshot}()$, it repeatedly reads
the array $\SM[1..m]$ until it obtains an array value $sm[1..m]$
that does not change during $(m(n-1)+2)$ readings of  $\SM[1..m]$. When this
occurs, the invoking process returns the corresponding array value $sm[1..m]$.

Trivially, any write operation terminates. As far the snapshot operation
is concerned, it is easy to see that, if there is a time after which a
process executes  alone it terminates its snapshot operation, hence the
implementation is obstruction-free.

To show that it is non-blocking, let us assume that a process invokes
repeatedly $\REG[x].{\sf write}()$ (whatever $x$) followed by
$\REG.{\sf snapshot}()$  (as it is the case in the algorithms presented
in the paper).
An invocation of $\REG.{\sf snapshot}()$ can be prevented from terminating
only if processes issue permanently invocations of  ${\sf write}()$,
Let us assume that no  invocation of  $\REG.{\sf snapshot}()$ terminates.
This means that there are processes that permanently issue write operations.
But this contradicts the assumption that each processes alternates
invocations of $\REG[x].{\sf write}()$ (whatever $x$) and
$\REG.{\sf snapshot}()$. This is because, between  two writes  issued
by a same process, this process invoked  $\REG.{\sf snapshot}()$, and
consequently this snapshot invocation terminated.

As far the linearization of the operations
${\sf write}()$ and  ${\sf snapshot}()$ invoked by the processes is
concerned we have the following (this proof is from~\cite{GR07}).
Let us consider an invocation of  ${\sf snapshot}()$ that terminates.
It has seen $m(n-1)+2$ times the same vector $sm[1..m]$ in
the array $\SM[1..m]$.
Since a given pair $\langle ts,v\rangle$ can be written at most once
by a process,  it can be written at most $(n-1)$ times during a snapshot
(once by each process, except the one invoking the snapshot).
It follows that, among the  $m(n-1)+2$ times where the same vector $sm[1..m]$
was read from $\SM[1..m]$, there are least two consecutive reads during which
no process wrote a register. The snapshot invocation is consequently
linearized after the first of these two reads.

%=========================================================================
\section{All Correct Processes Decide if One Process Decides}
\label{annex:one-more-register}
This appendix shows that, by adding one MWMR register, the consensus
termination property can be strengthened. More precisely, we have then
the additional termination property (where OA stands for ``One-All'').
\begin{itemize}
\vspace{-0.2cm}
\item OA-termination.
If a process decides, all correct processes decide.
\end{itemize}

Let $\DEC$ be the additional register, initialized to the default value
$\bot$.  The extended  algorithm is the one described in
Figure~\ref{fig:algorithm-OB-consensus} with only two modifications.

\begin{itemize}
\vspace{-0.2cm}
\item The first modification is the addition of the new line\\
\centerline{ {\bf if} $(\DEC\neq\bot)$
                {\bf then} ${\sf return}(\val)$ {\bf end if} }\\
between  line~\ref{Z-01} and line~\ref{Z-02}.
Each time it enters the repeat loop, a process first checks
if a value was previously decided. If it is the case, it decides it.
\vspace{-0.2cm}
\item The first modification is the addition, at line~\ref{Z-04},
 of the statement  ``$\DEC \leftarrow \val $'', just before
the statement ``${\sf return}(\val)$.
When a process is about to decide, it first writes the decided value in
the MWMR atomic register $\DEC$.
\end{itemize}

\begin{theorem}
\label{theo:OA-algorithm}
The extended  algorithm solves the obstruction-free consensus problem
satisfying the additional  {\em OA}-termination property, with $(n+1)$
underlying {\em MWMR} atomic registers.
\end{theorem}

\begin{proofT}
The proof follows directly from the proof of the base algorithm
of Figure~\ref{fig:algorithm-OB-consensus} (OB-termination and SV-termination)
and the fact that no process can block while executing the repeat loop
(hence OB-termination $\Rightarrow$ OA-termination).
\renewcommand{\toto}{theo:OA-algorithm}
\end{proofT}
%======================================================================

\end{document}